\documentclass{aa}

\usepackage[T1]{fontenc}
\usepackage[latin1]{inputenc}
\usepackage{wasysym}
\usepackage{graphicx}
\usepackage{natbib}
\usepackage{txfonts}

\newcommand{\eq}[1]{\begin{equation}  #1 \end{equation}}

\newcommand{\eqa}[1]{\begin{eqnarray}   #1 \end{eqnarray}}

\newcommand{\br}[1]{\left( #1 \right)}
\newcommand{\bc}[1]{\left\{ #1 \right\}}
\newcommand{\bb}[1]{\left[ #1 \right]}
\newcommand{\ba}[1]{\left\langle #1 \right\rangle}

\newcommand{\nn}{\nonumber}

\newcommand{\dd}{{\rm d}}
\newcommand{\expo}[1]{~{\rm e}^{ #1 }}
\newcommand{\vek}[1]{\mbox{\boldmath $#1$}}
\newcommand{\svek}[1]{\mbox{\boldmath \scriptsize $#1$}}  

\begin{document}

\title{The removal of shear-ellipticity correlations from the cosmic shear signal via nulling techniques}

\author{B. Joachimi \and P. Schneider}

\offprints{B. Joachimi,\\
    \email{joachimi@astro.uni-bonn.de}
}

\institute{Argelander-Institut f\"ur Astronomie (AIfA), Universit\"at Bonn, Auf dem H\"ugel 71, 53121 Bonn, Germany}

\date{Received 15 April 2008 / Accepted 22 May 2008}

\abstract{}
{To render cosmic shear an astronomical tool of high precision, it is essential to eliminate systematic effects upon its signal, one of the most significant ones being the intrinsic alignment of galaxies. The alignment in tidal fields that are created by the surrounding matter structure induces correlations between the intrinsic ellipticities of source galaxies, as well as correlations between the gravitational shear and the intrinsic ellipticity. While the former effect is restricted to physically close galaxy pairs and thus relatively easy to control, shear-ellipticity correlations occur for pairs at large separations. Because of the crudeness of current models of intrinsic alignment, we have developed a model-independent, purely geometrical method for removing the contamination of the cosmic shear signal by shear-ellipticity correlations.}
{We remove the contributions to a tomographic cosmic shear signal that may be subject to contamination by shear-ellipticity correlations, making use of the characteristic dependence of these correlations on redshift. By introducing an appropriately chosen weight function to the lensing efficiency that nulls signals stemming from certain distances, new second-order measures of cosmic shear can be constructed that are free from intrinsic alignment. We present three approaches to determining such weight functions, optimized with respect to the amount of information the weighting preserves. After generalizing the construction of weight functions, the loss of information induced by this nulling technique and the subsequent degradation of constraints on cosmological parameters is quantified in a likelihood analysis.}
{For constructing optimal weight functions, good agreement is achieved between all approaches considered. In particular, a simplified analytical ansatz is shown to approximate the numerical results closely, significantly lowering computational efforts. For a survey divided into 20 redshift bins, we find that the area of credible regions increases by $20\,\%$ up to about $50\,\%$ after the application of nulling, depending on the cosmological parameters considered. We demonstrate that, due to the optimization of the weight functions, nearly all information is contained in a small subset of the new second-order measures. The use of a significantly smaller number of redshift bins than 20 for the nulling considerably degrades parameter constraints under conservative assumptions, emphasizing the need for detailed redshift information.}
{}


\maketitle

\section{Introduction}

Cosmic shear, the gravitational lensing of distant galaxies by the intervening large-scale matter structure, provides a unique tool for cosmology, enabling cosmological parameters to be constrained based on only a few physical assumptions. Detected for the first time at the turn of the century by \citet{bacon00}, \citet{kaiser00}, \citet{vwaer00}, and \citet{wittman00}, cosmic shear observations have nowadays reached statistical errors on cosmological parameters compatible to those of other established methods \citep[see e.g.][]{jarvis06,hoekstra06,semboloni06,hetterscheidt07,benjamin07,fu07}. Lensing delivers information complementary to CMB anisotropies, Type Ia supernovae, or galaxy redshift surveys \citep[e.g.][]{spergel07,hu02b}, thereby playing a crucial role in the development of precision cosmology. In particular, cosmic shear surveys of the near future are currently considered to be among the most promising probes of dark energy and its parameters \citep[see][]{hu02,huterer02,albrecht06,peacock06}.

One of the major issues in rendering cosmic shear a tool of precision cosmology is the identification and removal of potential systematic effects that could have an influence on the shear signal if measured to high accuracy. Apart from the intricacies of systematic errors in observations, one has to take approximations made in the theory into account. Among these are the Born approximation, the dropping of lens-lens coupling and the replacement of the reduced shear by the shear itself, all three effects being a few per cent in magnitude \citep{Schneider98,SvWM02}. Physical effects to be considered are for instance source-lens and source-redshift clustering, where in both cases the contamination does not exceed the per cent level \citep[see][]{SvWM02}.

A potentially more serious modification of the cosmic shear signal is caused by correlations of intrinsic galaxy ellipticities, mimicking the effect of shear correlations. Consider a correlator of two galaxy ellipticities, which forms the base of all second-order cosmic shear measures. The (complex) measured ellipticity $\epsilon$ is the sum of the intrinsic ellipticity $\epsilon^{\rm s}$ of the galaxy and the gravitational shear $\gamma$. Writing the correlator for two galaxies $i$ and $j$ yields
\eq{
\label{eq:epscorrelators}
\ba{\epsilon_i \epsilon_j^*} = \ba{\gamma_i \gamma_j^*}+\ba{\epsilon_i^{\rm s} \epsilon_j^{{\rm s}*}}+\ba{\gamma_i \epsilon_j^{{\rm s}*}}+\ba{\epsilon_i^{\rm s} \gamma_j^*}\;.
}
For the following discussion assume that galaxy $i$ is situated closer to the observer, i.e. $z_i < z_j$.

The first term on the righthand side is the desired cosmic shear signal. The second term might be non-vanishing if the two galaxies were subject to the same tidal forces during their formation or evolution, for instance if they had formed in a single dark-matter halo. This can only happen if the galaxies are physically close, both on the sky and along the line of sight, so if $z_j - z_i \ll 1$. In the following we use the name \lq intrinsic ellipticity correlation\rq\ for this effect.

Various publications \citep[e.g.][]{croft00,catelan01,crittenden01,jing02} with both analytical and numerical approaches, e.g. via N-body simulations, have dealt with this possible contaminant of cosmic shear. They agree insofar as they predict intrinsic ellipticity correlations of 1 to $10\,\%$ of the lensing signal for surveys with median redshift around unity, whereas the effect may even dominate the cosmic shear signal for shallow surveys. This behavior is expected because a high median redshift corresponds to a broad redshift distribution of galaxies, which decreases the probability of finding two galaxies at the same redshift. \citet{brown02} detected intrinsic ellipticity correlations in the SuperCosmos field, which is shallow, so that all measured ellipticity correlations have to be purely intrinsic; see also \citet{mandelbaum06} for an analysis of SDSS data.

So far, only toy models exist for the intrinsic alignment of galaxies \citep[see e.g.][HS04 hereafter]{hirata04}, since little is known about the underlying processes of galaxy formation and evolution in their dark matter halos. However, as galaxies need to be physically close to mutually align, intrinsic ellipticity correlations are relatively easy to remove from the cosmic shear signal if redshift information is available. \citet{king02,king03} and \citet{heymans03} used different methods for downweighting galaxy pairs close in redshift, thereby reducing contamination by intrinsic alignment. Besides, intrinsic ellipticity correlations can generate B-modes \citep[HS04,][]{heymans06}, which could be used for its identification if alternative sources of a curl-component can be excluded.

The third term in (\ref{eq:epscorrelators}) describes the correlation between the intrinsic ellipticity of a background object and the shear signal of a foreground galaxy. As the lensing effect is generated by the matter structure between $z=0$ and $z=z_i$, there is no physical interaction between the matter acting as the lens and the galaxy at $z_j$, so that the correlation is expected to be identical to zero.

The last term can yield a contribution if the matter structure generates a tidal gravitational field in which a close-by galaxy aligns and by which a background object is gravitationally lensed. The foreground and background galaxies are found to be primarily oriented perpendicular to each other, resulting in an anti-correlation of the ellipticities. On small scales, a correlation might even stem from the alignment of a foreground galaxy with its own halo \citep{bridle07a}. We refer to this effect, first mentioned by HS04 as \lq shear-ellipticity correlation\rq\ (although a variety of names exists in the literature).

In contrast to the foregoing case, however, the removal of shear-ellipticity correlations is more complicated, as they are not restricted to physically close objects. On the contrary, their contribution increases for larger separations in redshift, becoming the dominant contamination for deep surveys (HS04). N-body simulations yield an upper limit of about 10$\,\%$ for shear-ellipticity correlations \citep{heymans06}, in agreement with the analytical approach of HS04. Such a contamination yields an underestimation (due to the anti-correlation) of $\sigma_8$ of about 5$\,\%$ and a bias on the parameters of the dark energy equation of state by up to 50$\,\%$ if acting together with intrinsic ellipticity correlations \citep[BK07 hereafter]{bridle07}. \citet{mandelbaum06} report the first observational verification of shear-ellipticity correlations in the Sloan Digital Sky Survey (SDSS) and estimated an upper limit of 20$\,\%$ contamination, while \citet{hirata07} find a best-fit intrinsic alignment model with data from SDSS and the 2SLAQ survey, which predicts $6.5\,\%$ contamination to the cosmic shear signal for a survey of similar galaxies.

The methods suggested for eliminating shear-ellipticity correlations from the cosmic shear signal mostly take advantage of the different redshift dependence compared to a pure lensing signal \citep[e.g. HS04,][BK07]{king05}, but they still partly rely on the uncertain models. Although \citet{heymans06} reported the detection of weak B-modes caused by shear-ellipticity correlations in their simulations, it is questionable whether this effect is strong enough to be unambiguously measured and therefore apt to remove the contamination. To provide a secure means of arriving at a data set free of shear-ellipticity correlations, we propose a purely geometrical method in the following, applicable to a cosmic shear survey with photometric redshift information. Making use of the characteristic dependence on redshift of these correlations, as already pointed out by HS04, we eliminate (\lq null\rq) the contributions to the cosmic shear signal that are possibly subject to a contamination by intrinsic alignment.

In Sect.$\,$\ref{sec:method} we introduce our notation, largely based on \citet{bartelmann01} and \citet{Schneider06}, and outline the basic elements of our method. We present two numerical and one analytical approach to determining weight functions, modifying the lensing efficiency to achieve the nulling of signals, in Sects.$\,$\ref{sec:numapproach} to \ref{sec:anaapproach}. Section \ref{sec:results} provides details on the implementation and the results for these approaches, while in Sect.$\,$\ref{sec:higherorder} the concept of weight functions is generalized to higher orders. The loss of information caused by the elimination of shear-ellipticity correlations and its consequences on constraints of cosmological parameters is discussed in Sect.$\,$\ref{sec:infoloss}. Finally, we conclude in Sect.$\,$\ref{sec:conclusions}.

\section{Method}
\label{sec:method}

The gravitational lensing of distant sources by the large-scale structure can be represented by the deflection of light in a single equivalent lens plane, neglecting lens-lens coupling and using the Born approximation. Accordingly, the three-dimensional mass distribution is projected onto this plane along the line of sight. Hence, for every comoving distance $\chi$ of a source, the dimensionless surface mass density or convergence $\kappa(\vek{\theta},\chi)$ can be obtained from the three-dimensional density contrast $\delta$ via
\eq{
\label{eq:kappawithchi}
\kappa(\vek{\theta},\chi) = \frac{3 H_0^2 \Omega_{\rm m}}{2 c^2} \int_0^\chi \dd \chi' \frac{f_{\rm k}(\chi') ~f_{\rm k}(\chi-\chi')}{f_{\rm k}(\chi)} ~\frac{\delta\br{f_{\rm k}(\chi') \vek{\theta},\chi'}}{a(\chi')}\;,
}
where $a(\chi)$ is the cosmic scale factor and
\eq{
\label{eq:fk2}
f_{\rm k}(\chi)=\left\{ \begin{array}{ll}
  \sin(\sqrt{K}\chi)/\sqrt{K}      &~~K>0\\
  \chi                             &~~K=0\\
  \sinh(\sqrt{-K}\chi)/\sqrt{-K}   &~~K<0\;.
\end{array} \right.
}
For more details see e.g. \citet{Schneider06}. In the following, the Universe is assumed for simplicity to be spatially flat, i.e. $K=0$. If photometric redshift information is available, several projections of $\kappa(\vek{\theta},\chi)$ can be constructed,
\eq{
\label{eq:projectedkappa}
\kappa^{(i)}(\vek{\theta}) = \int_0^{\chi_{\rm hor}} \dd \chi\; p^{(i)}(\chi)\; \kappa(\vek{\theta},\chi)\;,
}
where the upper limit of integration in (\ref{eq:projectedkappa}) is the comoving distance horizon. The weight function $p^{(i)}(\chi)$ denotes the normalized probability distribution of distances for a galaxy population $i$. In this work we assume that these distributions are disjunct redshift bins, i.e. we neglect the errors of photometric redshifts.

Consider now a tomographic shear signal from redshift bins $i$ and $j$, obtained via a second-order cosmic shear measure such as the correlation functions $\xi_\pm^{(ij)}(\theta)$, defined as usually in the context of weak gravitational lensing \citep[e.g.][]{Schneider06}. The contamination by shear-ellipticity correlations is caused by the matter distribution within the redshift slice situated closer to the observer, see (\ref{eq:epscorrelators}) and the discussion thereafter. If one cannot rely on a model governing the statistical properties of this contamination, it is necessary to eliminate the contribution of the matter within the lower redshift bin to the cosmic shear signal completely because only then can one ensure that shear-ellipticity correlations are not present anymore. Thus, the weighting in the projection of the density contrast has to be modified appropriately. Instead of using the galaxy distance distribution as in (\ref{eq:projectedkappa}), we now write
\eq{
\label{eq:projectedkappanew}
\tilde{\kappa}^{(i)}(\vek{\theta}) := \int_0^{\chi_{\rm hor}} \dd \chi\; B^{(i)}(\chi)\; \kappa(\vek{\theta},\chi)\;,
}
where $B^{(i)}(\chi)$ denotes the new weight function, measured in units of inverse length like $p^{(i)}(\chi)$. After inserting (\ref{eq:kappawithchi}) into (\ref{eq:projectedkappanew}) and rearranging the integrations, the relation of $\tilde{\kappa}(\vek{\theta})$ to the density contrast reads 
\eq{
\tilde{\kappa}^{(i)}(\vek{\theta}) = \frac{3 H_0^2 \Omega_{\rm m}}{2 c^2} \int_0^{\chi_{\rm hor}} \dd \chi ~\frac{\tilde{g}^{(i)}(\chi) ~ \chi}{a(\chi)} ~\delta\br{\chi \vek{\theta},\chi}\;,
}
with the lensing efficiency
\eq{
\label{eq:modlenseff}
\tilde{g}^{(i)}(\chi) = \int_\chi^{\chi_{\rm hor}} \dd \chi'\, B^{(i)}(\chi')\, \br{1 - \frac{\chi}{\chi'}}\;.
}
The lensing efficiency determines the amplitude by which the density contrast at the corresponding distance contributes to the convergence. Thus, demanding that the matter structure at a comoving distance $\hat{\chi}_i$ does not yield any contribution to the cosmic shear signal is equivalent to the constraint
\eq{
\label{eq:zwangsbed}
\tilde{g}^{(i)}(\hat{\chi}_i) = \int_{\hat{\chi}_i}^{\chi_{\rm hor}} \dd \chi ~B^{(i)}(\chi)\, \br{1 - \frac{\hat{\chi}_i}{\chi}} = 0\;.
}

Equation (\ref{eq:zwangsbed}) ensures the \lq nulling\rq\ of the contribution of matter at $\hat{\chi}_i$ to the shear signal, and consequently of the shear-ellipticity correlations, motivating the name \lq shear-ellipticity nulling\rq\ of this method. It makes use of the characteristic dependence on distance, $1-\hat{\chi}_i/\chi$, which corresponds to the ratio $D_{\rm ds}/D_{\rm s}$ of the angular diameter distance between lens and source and the one between observer and source. The nulling technique has been applied before in the context of cosmic shear by \citet{huterer05} who aimed at nulling small-scale information influenced by poorly determined baryonic physics. As $D_{\rm ds}/D_{\rm s}$ is a smooth function of both the lens and the source position, the integration in (\ref{eq:zwangsbed}) is rather insensitive to small changes in the distances of the mass distribution, acting as the lens. Consequently, the contributions from distances slightly smaller or larger than $\hat{\chi}_i$ are strongly downweighted by $B^{(i)}(\chi)$, too, provided that the weight function is also smooth. This leads to the near cancellation of the signal from the whole bin $i$ if $\hat{\chi}_i$ is chosen to be the distance corresponding to the center of the bin.

For reasons of simplicity, we assume in the following that the binning in redshift is equidistant with width $\Delta z$. However, the subsequent equations can also be formulated using variable bin sizes. We define a new second-order measure in terms of the tomography correlation functions as
\eq{
\label{eq:Xiapprox} 
\Xi^{(i)}_\pm(\theta) := \int_0^{\chi_{\rm hor}} \dd \chi ~B^{(i)}(\chi) ~\xi_\pm(z(\hat{\chi}_i),z(\chi),\theta)\;,
}
where the correlation functions $\xi_\pm(z_1,z_2,\theta)$ are -- formally -- evaluated at the exact, e.g. spectroscopically determined, redshifts $z_1$ and $z_2$. So far, $B^{(i)}(\chi)$ is only constrained for $\hat{\chi}_i < \chi \leq \chi_{\rm hor}$ by (\ref{eq:zwangsbed}). We set $B^{(i)}(\chi)\equiv0$ for $0 \leq \chi \leq \hat{\chi}_i$ because otherwise correlation functions with $z(\chi)<z(\hat{\chi}_i)$, which can be contaminated by intrinsic alignments at $z(\chi)$, would contribute to (\ref{eq:Xiapprox}). Through this, no information is discarded since the correlation functions are symmetric in their first and second arguments, so that, swapping the redshifts, they enter other measures, constructed like in (\ref{eq:Xiapprox}) but with smaller $\hat{\chi}_i$.

Now assume that $\xi_\pm(z(\hat{\chi}_i),z(\chi),\theta)$ is contaminated by shear-ellipticity correlations. Since $B^{(i)}(\chi)$ vanishes for $\chi<\hat{\chi}_i$, it is sufficient to consider $z(\chi)>z(\hat{\chi}_i)$ in the integral. Then intrinsic alignment has to be generated by the matter structure at $\hat{\chi}_i$; however, the contribution of matter at $\hat{\chi}_i$ to the convergence and subsequently to the shear entering the correlation function is eliminated by the weight function when chosen according to (\ref{eq:zwangsbed}), so that $\Xi^{(i)}_\pm(\theta)$ is free of shear-ellipticity correlations.

Since in practice only photometric redshift data will be available, we first transform (\ref{eq:Xiapprox}) to $z$ as the integration variable and subsequently approximate the integral as a Riemannian sum,
\eq{
\label{eq:Xiassum}
\Xi^{(i)}_\pm(\theta) \approx \sum_{j=1}^{N_z} B^{(i)}(\chi(z_j)) ~\xi^{(ij)}_\pm(\theta) ~\chi'(z_j) ~\Delta z\;,
}
where $N_z$ is the number of redshift bins and $\chi'(z)$ the derivative of the comoving distance with respect to $z$, calculated from
\eq{
\label{eq:wz}
\chi(z)=\frac{c}{H_0}\int^z_0 {\rm d}z' \bc{ \Omega_{\rm m} (1+z')^3 + \Omega_\Lambda }^{-1/2}\;.
}
The distance-redshift relation enters into many of the following equations since (\ref{eq:kappawithchi}) and related equations are formulated in terms of comoving distance, whereas the binning is performed in redshift, which is the observable. Henceforth, we refer to the first bin of the tomography correlation functions entering (\ref{eq:Xiassum}), i.e. the bin where the signal is nulled, as the \lq initial bin\rq. In (\ref{eq:Xiassum}) $z_j$ denotes the lower boundary redshift of bin $j$, implying that in the sum the correlation function $\xi^{(ij)}_\pm(\theta)$ is registered onto the lower boundary of bin $j$. To render (\ref{eq:Xiassum}) a good approximation, the redshift binning has to be sufficiently narrow. In addition, it is assumed that the redshift bins cover the total galaxy population almost completely, so that on discretizing (\ref{eq:Xiapprox}), the upper boundary of the integral can be reduced to the upper boundary of the highest redshift bin, denoted by $z_{\rm max}$.

For the sake of lower computational efforts, the further investigation of shear-ellipticity nulling is performed in Fourier space, by defining the new power spectrum
\eqa{
\nn
\Pi^{(i)}(\ell) &:=& \int_0^{\chi_{\rm hor}} \dd \chi ~B^{(i)}(\chi) ~P_\kappa(z(\hat{\chi}_i),z(\chi),\ell)\\
\label{eq:defPi}
&\approx& \sum_{j=1}^{N_z} B^{(i)}(\chi(z_j)) ~P_\kappa^{(ij)}(\ell) ~\chi'(z_j) ~\Delta z\;,
}
where $P_\kappa(z_1,z_2,\ell)$ denotes the convergence power spectrum for exactly known redshifts, and the angular frequency $\ell$ is the Fourier variable on the sky. The quantity $P_\kappa^{(ij)}(\ell)$ constitutes the corresponding tomographic measure, related to the three-dimensional power spectrum of matter density fluctuations $P_\delta$ via Limber's equation in Fourier space \citep{kaiser92},
\eq{
\label{eq:limber}
P_\kappa^{(ij)}(\ell) = \frac{9H_0^4 \Omega_{\rm m}^2}{4 c^4} \int^{\chi_{\rm hor}}_0 \dd \chi\; \frac{g^{(i)}(\chi)~g^{(j)}(\chi)}{a^2(\chi)} P_\delta \br{\frac{\ell}{\chi},\chi}\;,
}
where the lensing efficiency $g^{(i)}(\chi)$ in its original form, i.e. with $p^{(i)}(\chi)$ as weight, enters. However, since shear-ellipticity nulling requires detailed redshift information, $\Delta z \ll 1$. The weighted convergence in (\ref{eq:projectedkappa}) can then be written as $\kappa^{(i)}(\vek{\theta}) \approx \kappa(\vek{\theta},\hat{\chi}_i)$, or equivalently as $p^{(i)}(\chi) \approx \delta_{\rm D}(\chi -\hat{\chi}_i)$ to good approximation, where $\delta_{\rm D}$ is the Dirac delta distribution. In our case the dependence of the tomography power spectra on the probability distribution of galaxy distances is only marginal.

From the $N_z (N_z + 1)/2$ tomography power spectra, which are in principle available, only $N_z$ power spectra $\Pi^{(i)}(\ell)$ are constructed via (\ref{eq:defPi}). However, we determine more than just one weight function $B^{(i)}(\chi)$ per initial bin. Thus, as explained in Sects.$\,$\ref{sec:higherorder} and \ref{sec:infoloss}, the number of new power spectra will be equal to the number of input $P_\kappa^{(ij)}(\ell)$. Note that, after the explicit computation of weight functions, we show newly constructed power spectra and discuss their features; see Sect.$\,$\ref{sec:higherorder}.

As the weighting by $B^{(i)}(\chi)$ does not depend on angular scales, the considerations made above for $\Xi^{(i)}_\pm(\theta)$ also hold for $\Pi^{(i)}(\ell)$. The weight functions, once obtained, can be applied directly to measures accessible to observations, constructed in analogy to (\ref{eq:Xiassum}), by means of the transformations between convergence power spectra and real-space measures, as for instance given in \citet{SvWM02}. Hence, (\ref{eq:Xiapprox}) and (\ref{eq:defPi}) are related via
\eq{
\label{eq:FTxipi}
\Xi^{(i)}_\pm(\theta) = \int^\infty_0 \frac{\dd \ell ~\ell}{2 \pi} J_{0/4}(\ell \theta) ~\Pi^{(i)}(\ell)\;, 
}
where $J_\mu$ denotes the Bessel function of the first kind of order $\mu$.

In (\ref{eq:defPi}) it is sufficient to let the sum over redshift bins start only at $i+2$ because all other function values of $B^{(i)}(\chi)$ vanish. First, we demanded that $B^{(i)}(\chi)\equiv 0$ for $0 \leq \chi \leq \hat{\chi}_i$. In addition, to avoid contamination by intrinsic ellipticity correlations, we force the weight function to vanish within bin $i$ completely, implying that $B^{(i)}(\chi)$ departs from zero only from the upper boundary of bin $i$, i.e. from redshift $z_{i+1}$, onwards, so that the first non-vanishing value of the weight function in (\ref{eq:defPi}) is $B^{(i)}(\chi(z_{i+2}))$. 

\begin{figure}[h]
\centering
\includegraphics[scale=.38]{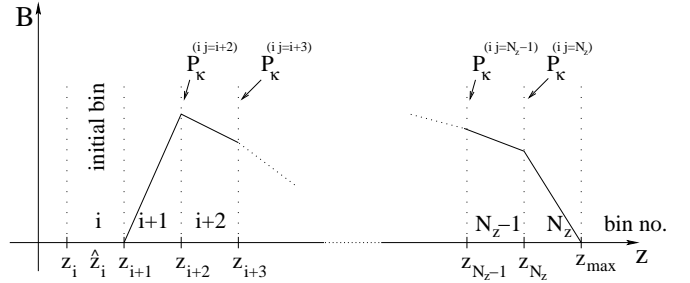}
\caption{Sketch illustrating the conventions made for the construction of weight functions. The redshift $z_j$ denotes the lower boundary of bin $j$; the tomography power spectrum $P^{(ij)}_\kappa(\ell)$ (or the corresponding real space measure) is registered onto this redshift in (\ref{eq:defPi}). With respect to these boundary redshifts, $B^{(i)}(\chi(z_{i+2}))$ is the first and $B^{(i)}(\chi(z_{N_z}))$ the last non-zero value of the weight function. The upper boundary redshift of the highest bin is denoted by $z_{\rm max}$. The initial bin is located between $z_i$ and $z_{i+1}$ with a central redshift of $\hat{z}_i$, so that $\hat{\chi}_i=\chi(\hat{z}_i)$.}
\label{fig:sketch}
\end{figure}

Consequently, $P^{(ij)}_\kappa(\ell)$ with $|j-i| \leq 1$, i.e. power spectra which auto-correlate redshift bins or cross-correlate adjacent bins are removed from the new measure (\ref{eq:defPi}), securing the downweighting of intrinsic ellipticity correlations. \citet{takada04b} demonstrate that errors on inferred parameters degrade by only about $10\,\%$ for 5 or more redshift bins if the auto-correlation power spectra are simply excluded from the analysis in shear tomography, which constitutes a simple and efficient method. Due to the narrow redshift binning our method requires, the contamination by intrinsic alignment may extend to neighboring bins (see e.g. BK07). Since we assume a choice of the redshift bins such that the number of galaxies with $z>z_{\rm max}$ is negligible, we set $B^{(i)}(\chi)\equiv0$ for $\chi \geq \chi(z_{\rm max})$. The conventions concerning the construction of weight functions made in this section are summarized in Fig.$\,$\ref{fig:sketch}.

\section{Determination of weight functions}
\label{sec:weightfunctions}

To show the feasibility of shear-ellipticity nulling, it is important to know to what extent and how efficient the weight function $B^{(i)}(\chi)$ can be constructed in practice. As a basic condition, $B^{(i)}(\chi)$ has to obey the constraint (\ref{eq:zwangsbed}). In addition, it is required that the weight function should be optimized in the sense that, when aiming at constraining cosmological parameters, the information content in the new power spectrum $\Pi^{(i)}(\ell)$ attains a maximum. This condition is quantified in terms of the Fisher matrix, which reads, when measuring the power spectrum in $N_\ell$ angular frequency bins, as
\eq{
\label{eq:fisherforpi}
F^{(i)}_{\mu \nu} = \sum_{\alpha,\,\beta=1}^{N_\ell} \left. \frac{\partial \Pi^{(i)}(\ell_\alpha)}{\partial p_\mu} \right|_{\rm f} \br{{C_\Pi^{(ii)}}^{-1}}_{\alpha \beta} \left. \frac{\partial \Pi^{(i)}(\ell_\beta)}{\partial p_\nu} \right|_{\rm f}\;.
}
The vector $\vek{p}$ is composed of the cosmological parameters that are considered. Its size $N_p$ implies the dimension $N_p \times N_p$ for the Fisher matrix. The index f is assigned to the power spectrum $\Pi^{(i)}(\ell)$, indicating that it is evaluated at the parameters of the fiducial cosmological model. The covariance $C^{(ii)}_\Pi$ measures the correlation of $\Pi^{(i)}$ between different angular frequency bins. From the definition (\ref{eq:defPi}) one finds
\eqa{
\label{eq:covPi}
\br{C_\Pi^{(ij)}}_{\alpha \beta} &=& \sum_{k=i+2,\,l=j+2}^{N_z} \br{C_P^{(ikjl)}}_{\alpha\beta}\; \\ \nn
&& \times\; B^{(i)}(\chi(z_k))\; B^{(j)}(\chi(z_l))\; \chi'(z_k)\; \chi'(z_l)\; \Delta z^2\;,
}
where the covariance of the convergence power spectra is given by (see \citealt{joachimi08} for details)
\eqa{
\nn
\br{C_P^{(ijkl)}}_{\alpha\beta} &=& \frac{2\pi}{A \ell_\alpha \Delta \ell_\alpha} \br{ \bar{P}^{(ik)}_\kappa(\ell_\alpha)\bar{P}^{(jl)}_\kappa(\ell_\alpha) + \bar{P}^{(il)}_\kappa(\ell_\alpha)\bar{P}^{(jk)}_\kappa(\ell_\alpha) }  \delta_{\alpha\beta}\\ 
\label{eq:covPtom}
&&\mbox{with}~~~  \bar{P}^{(kl)}_\kappa:=P^{(kl)}_\kappa+\delta_{kl} \frac{\sigma_\epsilon^2}{2\bar{n}^{(k)}}\;,
}
where $A$ denotes the size of the survey and $\Delta \ell$ the width of the angular frequency bins. The shape noise contribution is governed by the intrinsic galaxy ellipticity dispersion $\sigma_\epsilon$, i.e. $\sigma_\epsilon^2 = \langle \epsilon^{\rm s} \epsilon^{{\rm s}*}\rangle$ and by the mean number density of galaxies $\bar{n}^{(i)}$ in bin $i$. Note that for (\ref{eq:covPtom}) to be valid, we have assumed that the shear field is Gaussian and that the survey geometry is simple. Consequently, the power spectrum covariance is diagonal in angular frequency space, so that (\ref{eq:covPi}) is diagonal, too, rendering the inversion needed for (\ref{eq:fisherforpi}) trivial.

Besides, in (\ref{eq:covPtom}) it is assumed that any B-mode contribution to the power spectrum vanishes, which is expected for a pure lensing signal and which was already implicitly assumed throughout the preceding derivations. As a side remark, if there was a B-mode contribution due to shear-ellipticity correlations, the nulling technique could eliminate it as long as the B-mode signal has the same characteristic redshift dependence as the E-modes. However, though seen in simulations \citep{heymans06}, B-mode shear-ellipticity correlations are not expected from theory (HS04), so that the scaling with redshift is unknown.

As the actual quantity that is maximized, we choose the trace of the Fisher matrix because of its simple functional form, with the Fisher matrix elements entering ${\rm tr}(F^{(i)})$ linearly. Actually, the determinant of the Fisher matrix would be a more natural choice since its inverse is a measure of the volume of the error ellipsoid in parameter space, the inverse Fisher matrix being an estimate of the corresponding covariance matrix. However, due to the non-linearity of the determinant the numerical treatment is considerably less stable. As we demonstrate in Sect.$\,$\ref{sec:infoloss}, the trace fulfills the demand of concentrating the bulk of information into the power spectrum constructed out of the optimized weight function well enough. Exemplary, lower-dimensional calculations show that the difference in the form of the weight functions, using the trace or the determinant, is marginal.

Considering the continuous limit of $\Pi^{(i)}(\ell)$ in (\ref{eq:defPi}) for a moment, one arrives at a problem of variational calculus, determining the maximum of ${\rm tr}(F^{(i)})$ with respect to the function $B^{(i)}(\chi)$ under the constraint (\ref{eq:zwangsbed}). However, due to the binned redshift information, $B^{(i)}(\chi)$ only enters the equations in the form of discrete function values, turning the problem into a maximization with respect to these function values. The complete weight function is then constructed by either interpolation or in parameterized form. Plugging (\ref{eq:covPi}) and (\ref{eq:covPtom}) into (\ref{eq:fisherforpi}) yields
\eqa{
\label{eq:fulltrfisher}
{\rm tr}(F^{(i)}) &=& \frac{A}{2\pi} \sum_{\alpha=1}^{N_\ell} \ell_\alpha\; \Delta \ell_\alpha ~\Biggl[ ~\sum_{j,\,k=i+2}^{N_z} B(\chi(z_j))\; B(\chi(z_k))  \\ \nn
&& \hspace*{-.5cm} \times  ~\chi'(z_j)\; \chi'(z_k)\; \bc{ \bar{P}_\kappa^{(ij)}(\ell_\alpha) \bar{P}_\kappa^{(ik)}(\ell_\alpha) + \bar{P}_\kappa^{(ii)}(\ell_\alpha)  \bar{P}_\kappa^{(jk)}(\ell_\alpha)  } \Biggr]^{-1}\\ \nn
&& \hspace*{-1.5cm} \times \bb{\sum_{\mu=1}^{N_p} \br{ \sum_{j=i+2}^{N_z} \bc{ \frac{\partial P_\kappa^{(ij)}(\ell_\alpha)}{\partial p_\mu} \chi'(z_j) + P_\kappa^{(ij)}(\ell_\alpha) \frac{\partial \chi'(z_j)}{\partial p_\mu} } B(\chi(z_j)) }^2 }\;.
}
All power spectra and derivatives thereof are evaluated at the fiducial values of the cosmological model. Note that ${\rm tr}(F^{(i)})$ is independent of the overall amplitude of $B^{(i)}(\chi)$, as must be the case. The function values of $B^{(i)}(\chi)$ enter ${\rm tr}(F^{(i)})$ non-linearly, so that analytical progress is hampered in the general case. Both numerical approaches and analytical approximations are investigated in the following.

\subsection{Piecewise linear approach}
\label{sec:numapproach}

First, we consider a piecewise linear ansatz and write
\eq{
\label{eq:Bfem}
B^{(i)}(\chi(z)) := B_j + \frac{z-z_j}{\Delta z} \bc{B_{j+1} - B_j} ~~~\mbox{for}~~~ z \in \bb{z_j,z_{j+1}}\;,
}
where the notation $B_j \equiv B^{(i)}(\chi(z_j))$ was introduced for convenience. Moreover, we identify $B_{N_z+1} \equiv B^{(i)}(\chi(z_{\rm max}))=0$. The superscript $(i)$ of the weight function is dropped when using this shorthand notation. As before, $z_j$ denotes the lower boundary redshift of bin $j$, whereas we denote the central redshift of the initial bin, entering the constraint (\ref{eq:zwangsbed}), as $\hat{z}_i$. With the weight function in the form of (\ref{eq:Bfem}), the constraint reads
\eq{
\sum_{j=i+1}^{N_z} \int_{z_j}^{z_{j+1}} \!\!\!\!\!\dd z \bb{ B_j + \frac{z-z_j}{\Delta z} \bc{B_{j+1} - B_j} }\; \chi'(z) \br{1-\frac{\chi(\hat{z}_i)}{\chi(z)}} = 0\;.
}
Due to the conditions imposed on $B^{(i)}(\chi)$, as mentioned in Sect.$\,$\ref{sec:method}, $B_{i+1}=0$ holds, so that the constraint assumes the compact form
\eq{
\label{eq:Bfemcompact}
\sum_{j=i+2}^{N_z} B_j \br{I_j^0 - I_j^1 + I_{j-1}^1} = 0\;,
}
where the quantities
\eqa{
\label{eq:Bfemshorthand}
I_j^0 &=& \int_{z_j}^{z_{j+1}} \dd z\; \chi'(z) \br{1-\frac{\chi(\hat{z}_i)}{\chi(z)}}\\ \nn
&=& \chi(z_{j+1}) - \chi(z_j) - \chi(\hat{z}_i)\; \ln \br{\frac{\chi(z_{j+1})}{\chi(z_j)}}\;,\\ \nn 
I_j^1 &=& \int_{z_j}^{z_{j+1}} \dd z\; \frac{z - z_j}{\Delta z}\; \chi'(z) \br{1-\frac{\chi(\hat{z}_i)}{\chi(z)}}
}
were introduced. Fixing one of the function values via (\ref{eq:Bfemcompact}), the remaining $B_j$ are varied to obtain a maximum of the trace of the Fisher matrix, using a Nelder-Mead simplex algorithm, which is robust and does not require partial derivatives. However, this type of maximization routine easily gets stuck in local extrema, which in this case are caused by single outliers among the $B_j$. These are considered unrealistic since $B^{(i)}(\chi)$ is expected to be smooth, but the piecewise linear ansatz does not put any constraints on the derivatives of the weight function.

To avoid the false maxima due to outliers, we subtract the term
\eq{
\label{eq:punishterm}
U := \Lambda \sum_{j=i+1}^{N_z} \br{B_{j+1} - 2 B_j + B_{j-1}}^{2\eta}
}
from (\ref{eq:fulltrfisher}), summing the differences in slopes at the nodes, thereby disfavoring solutions with abrupt changes in the first derivative. The fudge factor $\Lambda>0$ can be chosen arbitrarily, while $\eta$ denotes a small, positive integer. The larger $\eta$, the more weight is given to large changes in slope in $U$, where we found a suitable value of $\eta=2$. If a value of $\Lambda$ is chosen such that $U$ and ${\rm tr}(F^{(i)})$ are roughly the same order of magnitude, the resulting weight functions are very smooth. By gradually lowering $\Lambda$, less smooth $B^{(i)}(\chi)$ with a more pronounced maximum are obtained. In case an outlier $B_j$ occurs, the initial values of the simplex algorithm are altered, until a stable solution with $\Lambda=0$ results.

The weight functions of this section are hardly susceptible to false maxima due to their smoothness, so that mostly we can set $\Lambda=0$ from the beginning. However, the procedure outlined here is necessary for the more oscillatory weight functions that will be computed in Sect.$\,$\ref{sec:higherorder}. Still, the final results presented in this work have been obtained with $\Lambda=0$ throughout. As can be seen from (\ref{eq:fulltrfisher}), the maximum of ${\rm tr}(F^{(i)})$ does not depend on the overall amplitude of $B^{(i)}(\chi)$, leading to a degeneracy in the maximized $B_j$, which will be lifted by a normalization, see Sect.$\,$\ref{sec:results}.

\subsection{Chebyshev series approach}
\label{sec:chebapproach}
The second numerical approach assumes that $B^{(i)}(\chi)$ is composed of a finite series of ansatz functions with a set of free parameters. In this case, we choose Chebyshev polynomials of the first kind $T_\mu$, which already lead to good approximations for low polynomial orders and yield evenly distributed errors. The weight function is expanded as
\eqa{
\label{eq:Bcheb}
B^{(i)}(\chi) &:=& \bc{\chi-\chi(z_{i+1})} \bc{\chi-\chi(z_{\rm max})}\\ \nn
&& \times \sum_{\mu=0}^{N_{\rm c}} b_\mu ~T_\mu\br{\frac{2\chi - \bc{\chi(z_{i+1})+\chi(z_{\rm max})}}{\chi(z_{\rm max})-\chi(z_{i+1})}}\;,
}
where the $b_\mu$ denote the $N_{\rm c}+1$ free coefficients. The argument of $T_\mu$ is chosen such that it takes on values in the interval $\bb{-1,1}$. Plugging this definition into (\ref{eq:zwangsbed}), one gets
\eq{
\label{eq:constraintcheb}
\sum_{\mu=0}^{N_{\rm c}} b_\mu Q^1_\mu = 0
}
with the definition
\eqa{
\label{eq:defQ}
Q^1_\mu &:=& \int_{z_{i+1}}^{z_{\rm max}} \dd z \bc{\chi(z)-\chi(z_{i+1})} \bc{\chi(z)-\chi(z_{\rm max})}\\ \nn
&& \times T_\mu\br{\frac{2\chi(z) - \bc{\chi(z_{i+1})+\chi(z_{\rm max})}}{\chi(z_{\rm max})-\chi(z_{i+1})}} \br{1-\frac{\chi(\hat{z}_i)}{\chi(z)}} \chi'(z)\;.
}

Again, one of the parameters is fixed by (\ref{eq:constraintcheb}), while $N_c$ parameters are used for the multi-dimensional maximization of the trace of the Fisher matrix with the simplex algorithm. If the number of free parameters is chosen to be more than about 5, the resulting weight functions are prone to significant oscillations, generated by the response of the ansatz polynomials to the steep rise in $B^{(i)}(\chi)$ for $\chi(z)-\chi(z_{i+1}) \ll 1$. These unphysical features, corresponding to shallow maxima in parameter space, are readily detected by visual inspection of the resulting weight functions and avoided by altering the -- in this approach low-dimensional -- set of initial values $b_\mu$. As for the piecewise linear ansatz, we observe the degeneracy in the parameters yielding the maximum due to the free scaling of the weight function.

\subsection{Simplified analytical approach}
\label{sec:anaapproach}

We elaborate on an analytical approach that is computationally fast and can provide an important consistency check for the preceding numerical methods. However, as stated above, the non-linearity hinders the analytical treatment of the full problem; instead, we confine ourselves in the following to considering a single angular frequency bin and only one element of the Fisher matrix, i.e. a single cosmological parameter. A vector notation is introduced as follows. 

Let the non-vanishing values of the weight function $B_j$, given in their shorthand notation of Sect.$\,$\ref{sec:numapproach}, form a vector $\vek{B}$. Note that in the vector notation we again drop the superscript $(i)$ since the initial bin that $\vek{B}$ refers to will be clear from the context. By defining another vector $\vek{f}$ with components
\eq{
\label{eq:defoff}
f_j := \br{1-\frac{\chi(\hat{z}_i)}{\chi(z_j)}} \;\chi'(z_j) ~~~~\mbox{for} ~~~j=i+2,\,..\, ,N_z\;,
}
the constraint (\ref{eq:zwangsbed}) simply turns in its discretized version into
\eq{
\label{eq:anaconstraint}
\br{\vek{B} \cdot \vek{f}}=0\;.
}
Note that the constant bin width $\Delta z$ is not included in $\vek{f}$, as it drops out when setting the discrete constraint expression to zero. The covariance of $\Pi^{(i)}$, given by (\ref{eq:covPi}), reduces to a scalar quantity due to the single angular frequency bin $\ell$ under consideration. Defining a matrix $\bar{C}$ with elements
\eq{
\bar{C}_{kl} := \br{C_P^{(ikil)}}_\ell\; \chi'(z_k)\; \chi'(z_l)\;,
}
the covariance can be written as 
\eq{
\label{eq:anacovpi}
C_\Pi^{(ii)} = \vek{B}^\tau \bar{C} \vek{B} ~{\Delta z}^2\;.
}
The Fisher matrix element, now indicated by a subscript $o$, reads as
\eq{
\label{eq:fisheranalyticalapproach}
F^{(i)}_o = \frac{\br{\vek{B} \cdot \vek{\rho}}^2}{\vek{B}^\tau \bar{C} \vek{B}}
}
with a further vector defined for convenience,
\eq{
\label{eq:defrho}
\rho_j := \frac{\partial P_\kappa^{(ij)}(\ell)}{\partial p} \chi'(z_j) + P_\kappa^{(ij)}(\ell) \frac{\partial \chi'(z_j)}{\partial p}\;.
}
By means of this vector, one is able to rewrite the derivative of the power spectrum with respect to the remaining cosmological parameter $p$ as \mbox{$\partial \Pi^{(i)}(\ell)/\partial p=\br{\vek{B} \cdot \vek{\rho}} \Delta z$}, which can be seen by taking the derivative of (\ref{eq:defPi}). The bin widths $\Delta z$ cancel in $F^{(i)}_o$, so that they do not need to appear in the definitions of $\bar{C}$ and $\vek{\rho}$. The constraint is incorporated by means of a Lagrange multiplier $\lambda$, leading to the quantity $G:=F^{(i)}_o+\lambda \br{\vek{B} \cdot \vek{f}}$, which is to be maximized with respect to the components of the vector $\vek{B}$. One obtains
\eq{
\nabla_B\; G = 2 \vek{\rho}\; \frac{\br{\vek{B} \cdot \vek{\rho}}}{\vek{B}^\tau \bar{C} \vek{B}} - 2\; \bar{C} \vek{B} \;\br{ \frac{\br{\vek{B} \cdot \vek{\rho}}}{\vek{B}^\tau \bar{C} \vek{B}} }^2 + \lambda \vek{f} = 0\;,
}
which can be formally solved for $\vek{B}$, resulting in
\eq{
\label{eq:Banaformal}
\vek{B} = {\cal N} \bar{C}^{-1} \bc{\vek{\rho}\; \frac{\vek{B}^\tau \bar{C} \vek{B}}{\br{\vek{B} \cdot \vek{\rho}}} + \frac{\lambda}{2}\; \vek{f}  \;\br{\frac{\vek{B}^\tau \bar{C} \vek{B}}{\br{\vek{B} \cdot \vek{\rho}}}}^2}\;,
}
where the free normalization ${\cal N}$ of $\vek{B}$ has been introduced.

\begin{figure*}[t]
\centering
\includegraphics[scale=.7,angle=270]{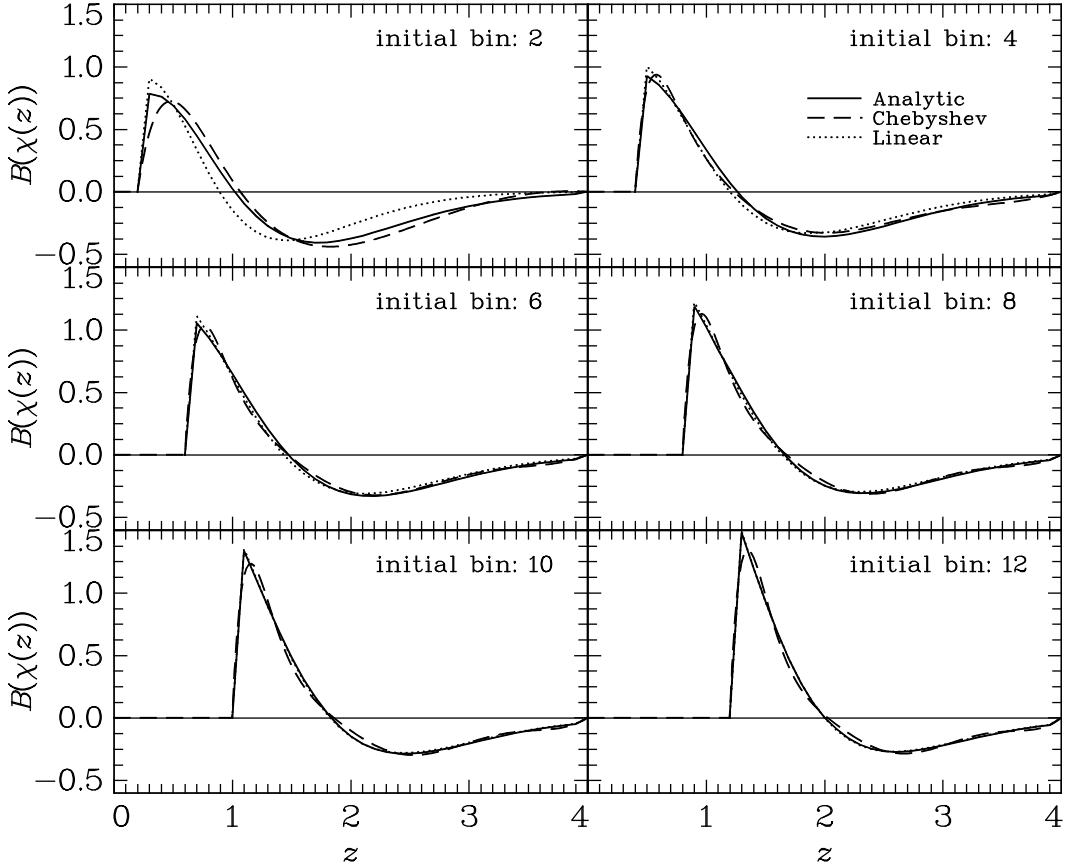}
\caption{Weight functions as a function of redshift for $N_z=40$. The initial bins are given in the respective panels. Plotted are the simplified analytical solution as solid curve, the Chebyshev series solution as dashed line, and the piecewise linear solution as dotted curve.}
\label{fig:wf_firstorder}
\end{figure*}

An overall scaling factor in $\vek{B}$ neither modifies the information content of $\Pi^{(i)}$ nor does it alter the constraint (\ref{eq:zwangsbed}), which illustrates that the conditions stated above do not fix the normalization. The formal solution is plugged into (\ref{eq:anaconstraint}), which is then solved for the Lagrange multiplier,
\eq{
\frac{\lambda}{2} = - ~\frac{\br{\vek{B} \cdot \vek{\rho}}}{\vek{B}^\tau \bar{C} \vek{B}} ~\frac{\vek{f}^\tau \bar{C}^{-1} \vek{\rho}}{\vek{f}^\tau \bar{C}^{-1} \vek{f}}\;.
}
 Replacing $\lambda/2$ in (\ref{eq:Banaformal}) subsequently yields
\eqa{
\label{eq:Bana}
\vek{B} &=& {\cal N} ~\frac{\vek{B}^\tau \bar{C} \vek{B}}{\br{\vek{B} \cdot \vek{\rho}}} \bc{\bar{C}^{-1} \vek{\rho} - \frac{\vek{f}^\tau \bar{C}^{-1} \vek{\rho}}{\vek{f}^\tau \bar{C}^{-1} \vek{f}}\; \bar{C}^{-1} \vek{f} }\\ \nn
&=& {\cal N'} \bc{\bar{C}^{-1} \vek{\rho} - \frac{\vek{f}^\tau \bar{C}^{-1} \vek{\rho}}{\vek{f}^\tau \bar{C}^{-1} \vek{f}}\; \bar{C}^{-1} \vek{f} }\;,
}
where the scalar quantity \mbox{$\br{\vek{B} \cdot \vek{\rho}}/\vek{B}^\tau \bar{C} \vek{B}$} was absorbed into the normalization ${\cal N'}$, so that now the righthand side depends no longer on $\vek{B}$. Optimized analytical weight functions can be calculated by means of (\ref{eq:Bana}), interpolating linearly between the values of the components of $\vek{B}$. In order to achieve results as close to the non-simplified, numerical approaches as possible, the employed element $F^{(i)}_o$ is chosen to be the diagonal element of the Fisher matrix that yields the largest contribution to the trace. Afterwards $\vek{B}$ is determined on a grid of $\ell$-values within the range considered in the numerical approaches, the solution vector resulting in the largest $F^{(i)}_o$ being taken as the \lq optimal\rq\ weight function. As long as this term, which is supposed to be the strongest contribution to the sum in (\ref{eq:fulltrfisher}), dominates the trace of the Fisher matrix, we expect this procedure to yield a reasonably good approximation to the numerical results.

\subsection{Setup \& results}
\label{sec:results}

To construct actual weight functions $B^{(i)}(\chi)$, a fictive tomographic cosmic shear survey with a comparatively large number of narrow redshift bins is needed. Due to the choice of $\Pi^{(i)}(\ell)$ as the quantity considered, the necessary input data comprises a set of tomographic power spectra, which are obtained for a $\Lambda$CDM universe with fiducial parameters $\Omega_{\rm m}=0.3$, $\Omega_{\rm \Lambda}=0.7$, and $H_0=100\, h_{100}\, {\rm km/s/Mpc}$ with $h_{100}=0.7$. The three-dimensional power spectrum of density fluctuations is specified by the primordial slope $n_{\rm s}=1$, the normalization $\sigma_8=0.9$ and the shape parameter $\Gamma$, calculated according to \citet{sugiyama95} with $\Omega_{\rm b}=0.04$. The linear power spectrum is given by the fit formula of \citet{bardeen86}, while the non-linear evolution is included via the prescription of \citet{smith03}. The tomography power spectra are then determined for $N_\ell=75$ logarithmic angular frequency bins between $\ell=50$ and $\ell=10^4$.

Furthermore, we specify survey properties that enter (\ref{eq:fulltrfisher}) via the power spectrum covariance (\ref{eq:covPtom}). For this a normalized galaxy redshift probability distribution
\eq{
\label{eq:redshiftdistribution}
p(z) \propto \br{\frac{z}{z_0}}^2 \exp \bc{ -\br{\frac{z}{z_0}}^\beta}
}
with $z_0=1.0$ and $\beta=1.5$ is assumed \citep{smail94}. The redshift distribution is cut off at $z_{\rm max}=4$, requiring a renormalization, which leads to the modified distribution $p_{\rm cut}(z)$. However, due to the large cut-off redshift the modification is marginal. The dependence of the covariance on the survey size $A$ is trivial, its value being irrelevant for the determination of the weight functions. For later calculations of likelihoods, we set $A$ to a fiducial size of $1\,{\rm deg}^2$. Moreover, we set the intrinsic ellipticity dispersion to $\sigma_\epsilon=0.4$ and choose a mean galaxy number density of $\bar{n}=30\,{\rm arcmin}^{-2}$. The bin-wise number densities are obtained by
\eq{
\bar{n}^{(i)} = \bar{n} \int_{z_i}^{z_{i+1}} \dd z ~p_{\rm cut}(z)\;.
}
The derivatives of the power spectra with respect to cosmological parameters in (\ref{eq:fulltrfisher}) are obtained via finite differencing, while $\partial \chi'(z_j)/\partial p_\mu$ is calculated from (\ref{eq:wz}) in analytical form.

To determine the Fisher information, we considered the set of cosmological parameters $\vek{p}=\br{\Omega_{\rm m},\Omega_\Lambda,\sigma_8,h_{100}}$. Aiming at smooth weight functions, we used a large number of redshift bins, i.e. $N_z=40$, corresponding to $\Delta z=0.1$. As mentioned in the foregoing section, the normalization of the weight functions is not yet fixed. To allow for direct comparison of the three approaches, we imposed the condition
\eq{
\label{eq:normalization}
\int_{\hat{\chi}_i}^{\chi_{\rm hor}} \dd \chi\; |B^{(i)}(\chi)|^2 = 1\;.
}
Furthermore, the free sign of $B^{(i)}(\chi)$ is chosen such that the weight function first assumes positive values when departing from zero at the upper boundary of the initial bin.

\begin{figure}[t]
\centering
\includegraphics[scale=.79,angle=270]{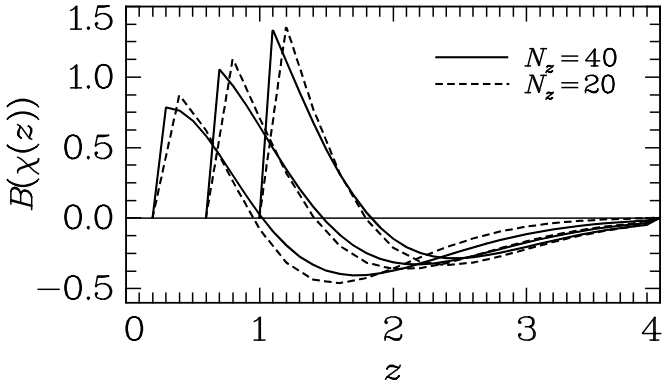}
\caption{Weight functions as a function of redshift for two different redshift binnings. Solid curves were obtained with $N_z=40$; dashed lines with $N_z=20$. Plotted are the simplified analytical solutions for initial bins 2, 6, and 10 in the case of 40 bins, and 1, 3, and 5 in the case of 20 bins.}
\label{fig:wf_compnzbins}
\end{figure}

In Fig.$\,$\ref{fig:wf_firstorder} the resulting weight functions of all three approaches under consideration are shown for varying initial bin $i$. All methods are in very good agreement; only in the upper left panel are larger deviations visible. A close inspection reveals that the weight functions constructed by means of the Chebyshev series suffer from slight oscillations that can rapidly increase in amplitude in some cases if $N_c$ is chosen too large. These can presumably be explained by the steep rise of $B^{(i)}(\chi)$ near the initial bin. Generally speaking, the agreement justifies the assumptions made in the different approaches; in particular, the results of the relatively crude approximations of the analytical approach are compatible with the numerical calculations, so that it is well-suited to further investigation. The weight functions have a zero-crossing, which is expected due to (\ref{eq:zwangsbed}), where the term $1-\hat{\chi}_i/\chi$ is non-negative throughout the integration interval. The largest weight is assigned to those redshifts that are located directly above the initial bin, respectively, because the efficiency of the lensing of a source in this range of redshifts by the mass distribution within the initial bin is low or, in other words, $D_{\rm ds}/D_{\rm s}$ is small, which decreases the contribution by shear-ellipticity correlations.

The division of a cosmic shear survey into 40 redshift bins is realistic in the near future; however, the bin size would not be chosen constant as in this study, but probably scale with $1+z$. Mainly for computational reasons, we reduce the number of redshift bins used in the following likelihood analysis to 20, a number achieved by some of the upcoming wide-field projects such as Pan-Starrs, KIDS/Viking, the Dark Energy Survey, LSST, or Euclid. We compare the form of the weight functions obtained above with an analogous set, determined for $N_z=20$. Figure \ref{fig:wf_compnzbins} illustrates for a sample of analytical solutions for $B^{(i)}(\chi)$ that the sets for both redshift binnings agree well. The higher density of sampling points in the case of $N_z=40$ enables a steeper rise of the weight functions at the upper boundary of the initial bin, leading to deviations in similar magnitude in the tail of $B^{(i)}(\chi)$. These results also suggest that the effects due to the discretization of the weight functions are negligible as long as the number of redshift bins is not chosen too small.

\subsection{Higher orders}
\label{sec:higherorder}

\begin{figure*}[t]    
\centering
\includegraphics[scale=.7,angle=270]{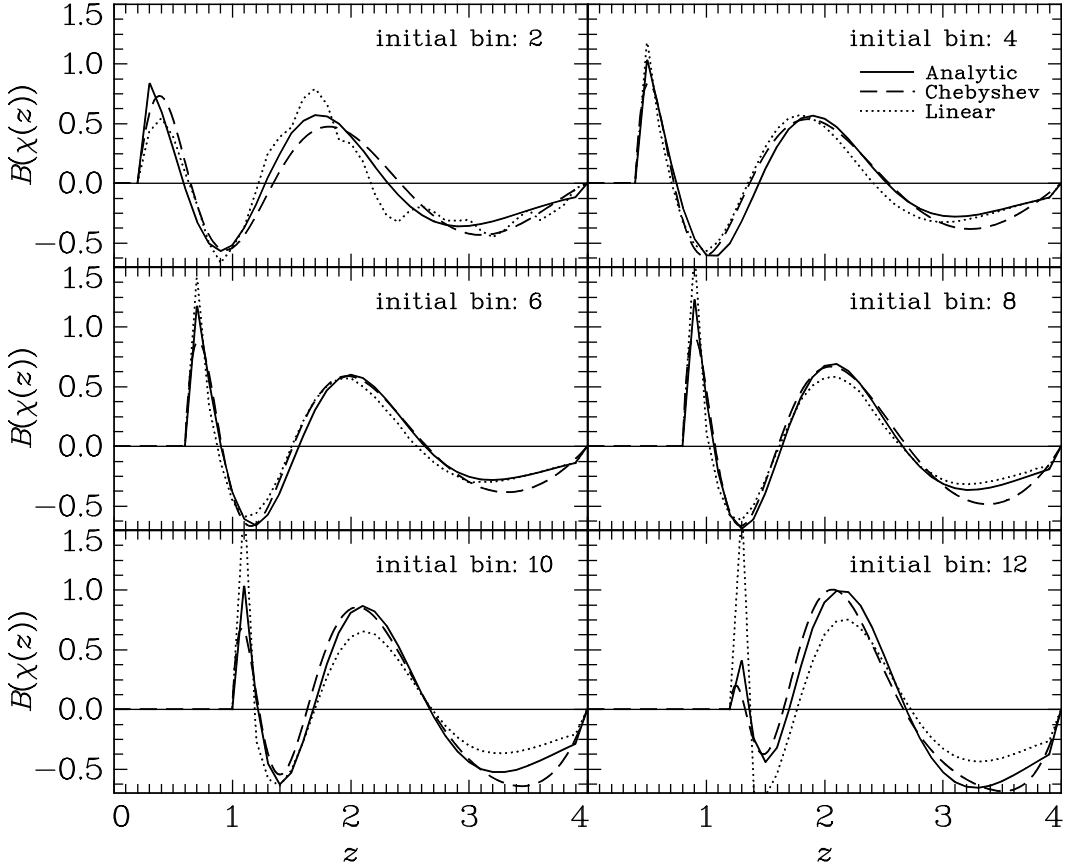}
\caption{Second-order weight functions as a function of redshift for $N_z=40$. The initial bins are given in the respective panels. The coding of the curves is the same as in Fig.$\,$\ref{fig:wf_firstorder}.}
\label{fig:wf_secondorder}
\end{figure*}

The power spectrum $\Pi^{(i)}(\ell)$, given by (\ref{eq:defPi}), is a linear combination of the convergence power spectra, with the weighting determined by the function $B^{(i)}(\chi)$, as calculated in the foregoing section. More such linear combinations can be constructed with differing weight functions that still obey the constraint equation, resulting in further power spectra free of shear-ellipticity correlations. If one retains the condition of maximizing the Fisher matrix and, in addition, demands that the weight functions should be orthogonal with respect to each other in a suitably defined sense, one arrives at higher-order measures that have the second-most, third-most, etc., information content.

Since in (\ref{eq:defPi}) the first term that yields a contribution is for $j=i+2$ (see also Fig.$\,$\ref{fig:sketch}), $N_z-i-1$ convergence power spectra are used to form $\Pi^{(i)}(\ell)$ in the implementation presented above. As a consequence, one is able to construct $N_z-i-2$ mutually orthogonal power spectra $\Pi^{(i)}(\ell)$ from this data set. The additional combination that could be built furthermore with linear independence from the set of convergence power spectra then necessarily violates (\ref{eq:zwangsbed}) and consequently contains shear-ellipticity correlations.

Denoting the order of the weight function by a subscript in square brackets, the condition of mutual orthogonality between weight functions of order $q$ and $r$ can be formulated as
\eq{
\label{eq:orthogonality}
\int_{\hat{\chi}_i}^{\chi(z_{\rm max})} \dd z ~B^{(i)}_{[q]}(\chi(z)) ~B^{(i)}_{[r]}(\chi(z)) ~w(z) = 0
}
for all orders $q > r$, where $w(z)$ is an arbitrary weight function. As far as the two numerical approaches are concerned, the higher order weight functions are obtained by fixing one further free parameter for every orthogonality condition. In the case of the piecewise linear ansatz, one obtains by plugging the ansatz functions (\ref{eq:Bfem}) into (\ref{eq:orthogonality})
\eqa{
\nn
\sum_{j=i+1}^{N_z} \int_{z_j}^{z_{j+1}} \dd z \bb{ B_{[q],j} + \frac{z-z_j}{\Delta z} \bc{B_{[q],j+1} - B_{[q],j}} } &&\\
\label{eq:ortholin}
&& \hspace*{-2cm} \times\; B^{(i)}_{[r]}(\chi(z))\; w(z) = 0\;
}
for every $r=1,\,..\,,q-1$. The lower order weight functions have been determined in advance and are known. Defining
\eqa{
I_{r,j}^2 &=& \int_{z_j}^{z_{j+1}} \dd z\; B^{(i)}_{[r]}(\chi(z))\; w(z)\;,\\ \nn
I_{r,j}^3 &=& \int_{z_j}^{z_{j+1}} \dd z\; \frac{z-z_j}{\Delta z}\; B^{(i)}_{[r]}(\chi(z))\; w(z)\;,
}
one can write (\ref{eq:ortholin}) in analogy to (\ref{eq:Bfemcompact}) as 
\eq{
\label{eq:ortholincompact}
\sum_{j=i+2}^{N_z} B_{[q],j} \br{I_{r,j}^2 - I_{r,j}^3 + I_{r,j-1}^3} = 0\;.
}
The Chebyshev approach yields, inserting (\ref{eq:Bcheb}) into (\ref{eq:orthogonality}),
\eq{
\label{eq:orthocheb}
\sum_{\mu=0}^{N_{\rm c}} b_\mu Q^2_{r,\mu} = 0\;,
}
again for $r=1,\,..\,,q-1$, where we have defined
\eqa{
\label{eq:defQortho}
Q^2_{r,\mu} &:=& \int_{z_{i+1}}^{z_{\rm max}} \dd z \bc{\chi(z)-\chi(z_{i+1})} \bc{\chi(z)-\chi(z_{\rm max})}\\ \nn
&& \times T_\mu\br{\frac{2\chi(z) - \bc{\chi(z_{i+1})+\chi(z_{\rm max})}}{\chi(z_{\rm max})-\chi(z_{i+1})}} B^{(i)}_{[r]}(\chi(z))\; w(z)\;.
}

Besides, higher order weight functions can be constructed with the analytical ansatz, again considering only a single angular frequency bin and solely one component of the Fisher matrix, optimizing it subsequently as outlined in Sect.$\,$\ref{sec:anaapproach}. To clarify the notation, we rewrite the result for the first-order weight function (\ref{eq:Bana}) as
\eq{
\label{eq:firstorderrewritten}
\vek{B}_{[1]} = {\cal N'} \bar{C}^{-1} \vek{\rho}^{\bc{1}} ~~~~\mbox{with}~~~~ \vek{\rho}^{\bc{1}} := \vek{\rho} - \frac{\vek{f}^\tau \bar{C}^{-1} \vek{\rho}}{\vek{f}^\tau \bar{C}^{-1} \vek{f}}\; \vek{f}\;.
}
As in Sect.$\,$\ref{sec:anaapproach} we perform the derivation for an initial bin $i$, where the index does not explicitly appear in the formulae, but enters the quantities $\vek{f}$, $\bar{C}$, and $\vek{\rho}$. In this context the condition of orthogonality can be implemented as
\eq{
\label{eq:orthogonalityana}
\br{\vek{B}_{[q]} \cdot \vek{\widetilde{B}}_{[r]}}=0
}
for all orders $q > r$, where $\widetilde{B}_j:=B^{(i)}(\chi(z_j)) w(z_j)$ was defined. This condition is incorporated into the maximization by more Lagrange multipliers, the expression to be maximized for order $q$ turning into
\eq{
G_{[q]} = F^{(i)}_o + \lambda_{[q]} \br{\vek{B}_{[q]} \cdot \vek{f}} + \sum_{r=1}^{q-1} \mu^r_{[q]} \br{\vek{B}_{[q]} \cdot \vek{\widetilde{B}}_{[r]}}\;,
}
where the $\mu^r_{[q]}$ are the Lagrange multipliers for the respective orthogonality conditions. The Fisher matrix element $F^{(i)}_o$ is still given by (\ref{eq:fisheranalyticalapproach}), now with weight functions $\vek{B}_{[q]}$. After taking the gradient with respect to the components of $\vek{B}_{[q]}$ in analogy to the first-order calculation, one arrives at the formal solution
\eqa{
\label{eq:bhigherformal}
\vek{B}_{[q]} &=& {\cal N} \bar{C}^{-1} \Biggl\{ \vek{\rho}\; \frac{\vek{B}_{[q]}^\tau \bar{C} \vek{B}_{[q]}}{\br{\vek{B}_{[q]} \cdot \vek{\rho}}} + \frac{\lambda_{[q]}}{2}\; \vek{f}\;  \br{\frac{\vek{B}_{[q]}^\tau \bar{C} \vek{B}_{[q]}}{\br{\vek{B}_{[q]} \cdot \vek{\rho}}}}^2\\ \nn
&& \hspace*{2.5cm} + \sum_{r=1}^{q-1}  \frac{\mu^r_{[q]}}{2}\; \vek{\widetilde{B}}_{[r]}\; \br{\frac{\vek{B}_{[q]}^\tau \bar{C} \vek{B}_{[q]}}{\br{\vek{B}_{[q]} \cdot \vek{\rho}}}}^2 \Biggr\}\;.
} 
The Lagrange multipliers are successively replaced by inserting this solution into the corresponding constraint equations. As a first step, from $\br{\vek{f} \cdot \vek{B}_{[q]}}=0$ one obtains
\eq{
\frac{\lambda_{[q]}}{2} = - \frac{\br{\vek{B}_{[q]} \cdot \vek{\rho}}}{\vek{B}_{[q]}^\tau \bar{C} \vek{B}_{[q]}} ~\frac{\vek{f}^\tau \bar{C}^{-1} \vek{\rho}}{\vek{f}^\tau \bar{C}^{-1} \vek{f}} - \sum_{r=1}^{q-1}  \frac{\mu^r_{[q]}}{2} ~\frac{\vek{f}^\tau \bar{C}^{-1} \vek{\widetilde{B}}_{[r]}}{\vek{f}^\tau \bar{C}^{-1} \vek{f}}\;.
}
Plugging in this expression, (\ref{eq:bhigherformal}) turns into
\eq{
\label{eq:banahostep}
\vek{B}_{[q]} = {\cal N'} \bar{C}^{-1} \bc{\vek{\rho}^{\bc{1}} + \sum_{r=1}^{q-1}  \frac{\mu^r_{[q]}}{2}\; \vek{\widetilde{B}}_{[r]}^{\bc{1}} \frac{\vek{B}_{[q]}^\tau \bar{C} \vek{B}_{[q]}}{\br{\vek{B}_{[q]} \cdot \vek{\rho}}}}
}
 with the definition
\eq{
\label{eq:initialrecursion}
\vek{x}^{\bc{1}} := \vek{x} - \frac{\vek{f}^\tau \bar{C}^{-1} \vek{x}}{\vek{f}^\tau \bar{C}^{-1} \vek{f}}\; \vek{f}\;,
}
where $\vek{x} \in \bc{\vek{\rho},\vek{B}_{[q]}}$. Again, multiplicative scalars have been absorbed into the normalization. In a similar manner, inserting (\ref{eq:banahostep}) into (\ref{eq:orthogonalityana}) for $r=1$ leads to
\eq{ 
\vek{B}_{[q]} = {\cal N'} \bar{C}^{-1} \bc{\vek{\rho}^{\bc{2}} + \sum_{r=2}^{q-1}  \frac{\mu^r_{[q]}}{2}\; \vek{\widetilde{B}}_{[r]}^{\bc{2}}\; \frac{\vek{B}_{[q]}^\tau \bar{C} \vek{B}_{[q]}}{\br{\vek{B}_{[q]} \cdot \vek{\rho}}}}\;,
}
where we set
\eq{
\label{eq:recursionstep2}
\vek{x}^{\bc{2}} := \vek{x}^{\bc{1}} - \frac{\vek{\widetilde{B}}_{[1]}^\tau \bar{C}^{-1} \vek{x}^{\bc{1}}}{\vek{\widetilde{B}}_{[1]}^\tau \bar{C}^{-1} \vek{\widetilde{B}}_{[1]}^{\bc{1}}}\; \vek{\widetilde{B}}_{[1]}^{\bc{1}}\;.
}
If one continues likewise for the remaining constraint equations, one obtains in accordance with (\ref{eq:firstorderrewritten}) the compact result
\eq{
\label{eq:Bhigherordergeneral}
\vek{B}_{[q]} = {\cal N} \bar{C}^{-1} \vek{\rho}^{\bc{q}}
}
for all orders $q$, where the redefined normalization is denoted by just ${\cal N}$ again. Here we made use of the recursion relation
\eq{
\vek{x}^{\bc{r}} = \vek{x}^{\bc{r-1}} - \frac{\vek{\widetilde{B}}_{[r-1]}^\tau \bar{C}^{-1} \vek{x}^{\bc{r-1}}}{\vek{\widetilde{B}}_{[r-1]}^\tau \bar{C}^{-1} \vek{\widetilde{B}}_{[r-1]}^{\bc{r-1}}}\; \vek{\widetilde{B}}_{[r-1]}^{\bc{r-1}}\;,
}
supplemented by the initial step (\ref{eq:initialrecursion}). Consequently, a recursion relation for the vectors $\vek{B}_{[q]}$, corresponding to the higher order weight functions, can be derived, which reads
\eq{
\label{eq:Brecursive}
\vek{B}_{[q]} = {\cal N} \bc{\vek{B}_{[q-1]} - \frac{\vek{\widetilde{B}}_{[q-1]}^\tau ~\vek{B}_{[q-1]}}{\vek{\widetilde{B}}_{[q-1]}^\tau \bar{C}^{-1} \vek{\widetilde{B}}_{[q-1]}^{\bc{q-1}}}\; \bar{C}^{-1} \vek{\widetilde{B}}_{[q-1]}^{\bc{q-1}} } 
}
for $q \geq 2$, the vector $\vek{B}_{[1]}$ as the starting point for this recursion being given by (\ref{eq:firstorderrewritten}).

\begin{figure}[t]
\centering
\includegraphics[scale=.79]{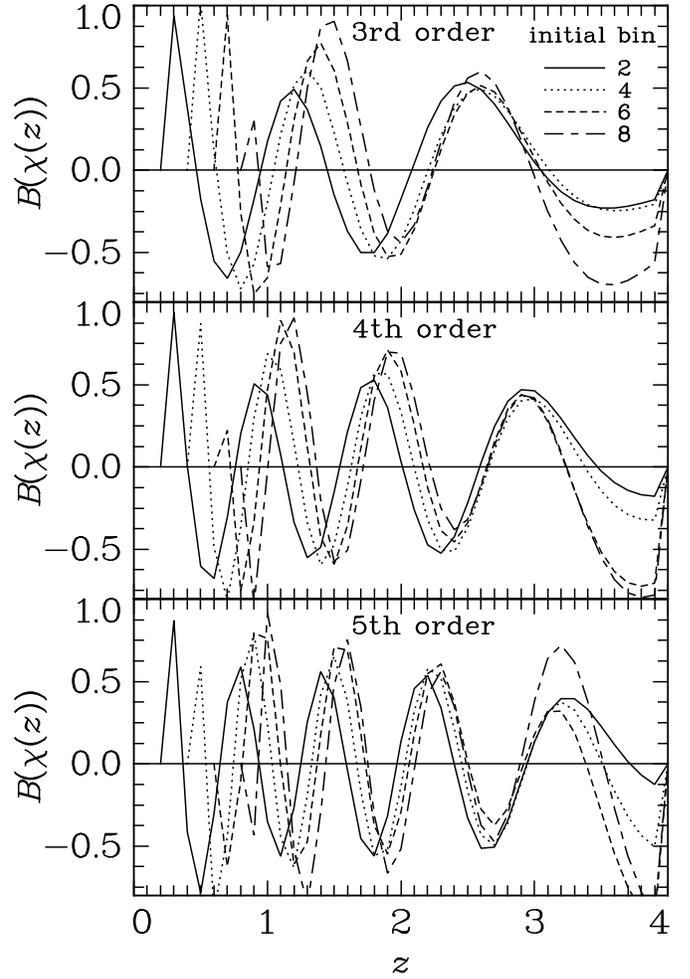}
\caption{Higher-order weight functions as a function of redshift for $N_z=40$. From top to bottom the third to fifth order analytical solutions are shown. Solid curves correspond to initial bin no.$\,$2, dotted curves to initial bin no.$\,$4, short-dashed curves to initial bin no.$\,$6, and long-dashed curves to initial bin no.$\,$8.}
\label{fig:wf_highorder}
\end{figure}

The weight function $w(z)$ could for instance be chosen, such that it scales with the redshift probability distribution $p_{\rm cut}(z)$, assigning a larger weight to well-sampled redshift ranges. However, we set $w(z)\equiv 1$ in the following for reasons of simplicity. In addition, (\ref{eq:orthogonalityana}) then turns into an orthogonality relation also for the vectors $\vek{B}_{[q]}$.

In Fig.$\,$\ref{fig:wf_secondorder} the results for second-order weight functions $B^{(i)}_{[2]}(\chi)$ are plotted, for all three methods considered in this work and using the same setup as described in Sect.$\,$\ref{sec:results}. Apart from slight numerical instabilities, as can be seen for the linear approach in the upper left panel, and the differing response to sharp peaks in the weight functions, most prominent in the lower right panel, the curves largely agree. Again, the simplified analytical ansatz proves to be compatible, being computationally advantageous to a large extent due to its recursive form (\ref{eq:Brecursive}). For all curves the number of zeros has increased by two compared to the first-order results, a trend that continues for higher orders. A sample of analytical solutions for third- to fifth-order weight functions is given in Fig.$\,$\ref{fig:wf_highorder}.

\begin{figure}[t]
\centering
\includegraphics[scale=.54]{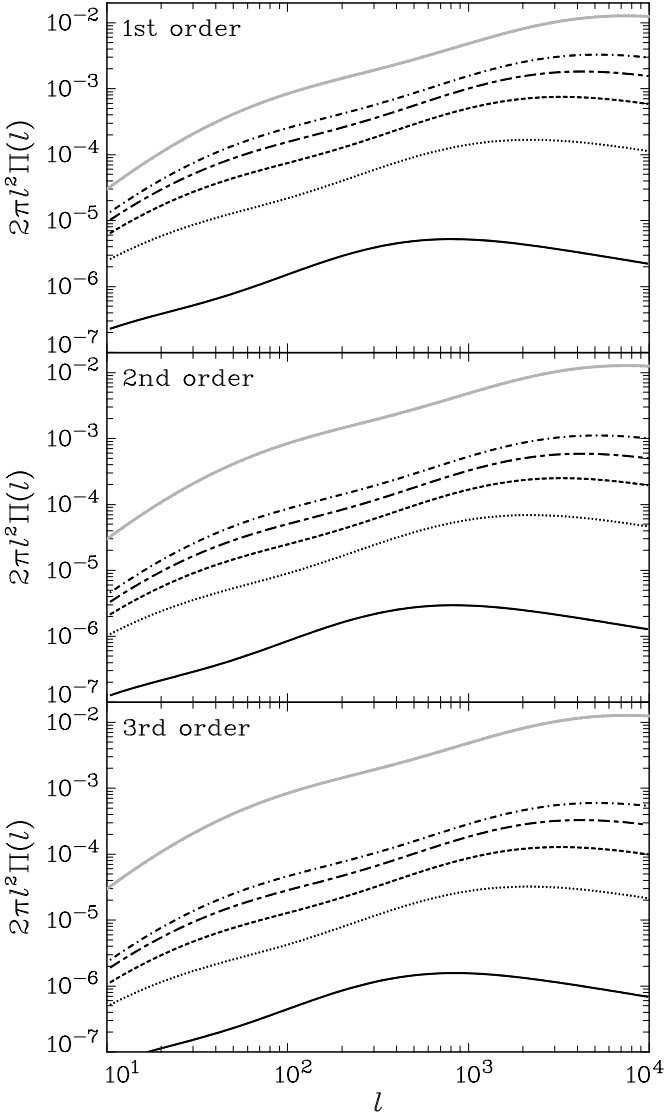}
\caption{
New power spectra as a function of angular frequency, making use of analytically determined weight functions. The power spectra  $\Pi^{(i)}_{[q]}(\ell)$ are given as black curves, their order $q$ ranging from 1 in the top panel to 3 in the bottom panel. Within each panel the power spectra for initial bins $i=1,\,..\,,5$ are plotted in the following sequence of line types: solid, dotted, dashed, chain-dashed, and dot-dashed. In addition, the convergence power spectrum $P_\kappa(\ell)$, integrated over the redshift distribution (\ref{eq:redshiftdistribution}), is shown for reference as gray curve.
}
\label{fig:wf_piplot}
\end{figure}

Adopting the notation introduced for the weight functions, the new power spectra can be generalized to higher orders as
\eq{
\label{eq:defPigen}
\Pi^{(i)}_{[q]}(\ell) \approx \sum_{j=1}^{N_z} B^{(i)}_{[q]}(\chi(z_j)) ~P_\kappa^{(ij)}(\ell) ~\chi'(z_j) ~\Delta z
}
for initial bin $i$ and order $q$. Employing analytically determined weight functions, we compute these power spectra for different initial bins and orders, the resulting graphs shown in Fig.$\,$\ref{fig:wf_piplot}. For reference the convergence power spectrum $P_\kappa(\ell)$, integrated over the full redshift distribution as given by (\ref{eq:redshiftdistribution}), is plotted in addition. It is important to note that $\Pi^{(i)}_{[q]}(\ell)$ and $P_\kappa(\ell)$ can only be compared with difficulty in terms of the overall amplitude, since for the newly constructed power spectra, the amplitude can be chosen arbitrarily due to the free normalization of the $B^{(i)}_{[q]}(\chi)$. In Fig.$\,$\ref{fig:wf_piplot} it is fixed by (\ref{eq:normalization}), so that the weights are of order unity. Therefore it is evident that the $\Pi^{(i)}_{[q]}(\ell)$ have considerably lower amplitude than the reference power spectrum since to obtain the former quantities, power spectrum signals are partially subtracted.

Concerning shape, the $\Pi^{(i)}_{[q]}(\ell)$ show a largely similar behavior with respect to the convergence power spectrum, the latter peaking at higher values of $\ell$. This can be understood by taking into account that the tomography power spectrum with the smallest difference between bins $i$ and $j$ contributes most to the respective $\Pi^{(i)}_{[q]}(\ell)$, as can be concluded from the form of the first-order weight functions, its pronounced peak being located just above the initial bin, see Fig.$\,$\ref{fig:wf_firstorder}. Thus, the new power spectra receive their signal preferentially from less distant galaxies, so that they probe smaller physical separations for fixed angular scale or $\ell$, respectively. On small scales non-linear structure evolution sets in, enhancing the signal. Consequently, the characteristic bump caused by non-linearity is visible for smaller $\ell$, i.e. larger angles, in the new power spectra in comparison with $P_\kappa(\ell)$, leading also to the shift of the peak.

\section{Information loss}
\label{sec:infoloss}

By eliminating contributions to the cosmic shear signal at certain distances from the observer, one necessarily reduces the information content of the data set, so that the desired constraints on cosmological parameters are less stringent. Thus, to judge the practical value of the nulling technique, we are going to quantify the accuracy with which cosmological parameters can be determined by the newly constructed power spectra (\ref{eq:defPi}). 
If one considers the set of $N_z-i-1$ tomography power spectra used to construct $\Pi^{(i)}_{[q]}(\ell)$ as the components of a data vector, then shear-ellipticity nulling is equivalent to a rotation of this vector such that all but one component of the resulting vector are free of shear-ellipticity correlations. The \lq cleaned\rq\ components correspond to the \mbox{$N_z-i-2$} new power spectra (\ref{eq:defPigen}) for $q=1,\,..\,,N_z-i-2$, whereas the last component must contain a weight function that is collinear to $\vek{f}$, or simply
\eq{
\Pi^{(i)}_{[N_z-i-1]}(\ell) \approx \sum_{j=1}^{N_z} P_\kappa^{(ij)}(\ell) \br{1-\frac{\chi(\hat{z}_i)}{\chi(z_j)}} ~\chi'(z_j) ~\Delta z\;;
}
i.e., $\vek{f}$ itself is chosen as the weight, see (\ref{eq:defoff}). The vector rotation mentioned above is invertible, so that, illustratively, it is obvious that using the new, full data vector instead of the one containing the convergence power spectra for the data analysis, one should obtain the same results. This statement is equivalent to the Fisher matrix, as a measure of the information content, being invariant under such orthogonal transformations of the data vector \citep[see][]{tegmark97}.

\begin{figure*}[t]
\centering
\includegraphics[scale=.9,angle=270]{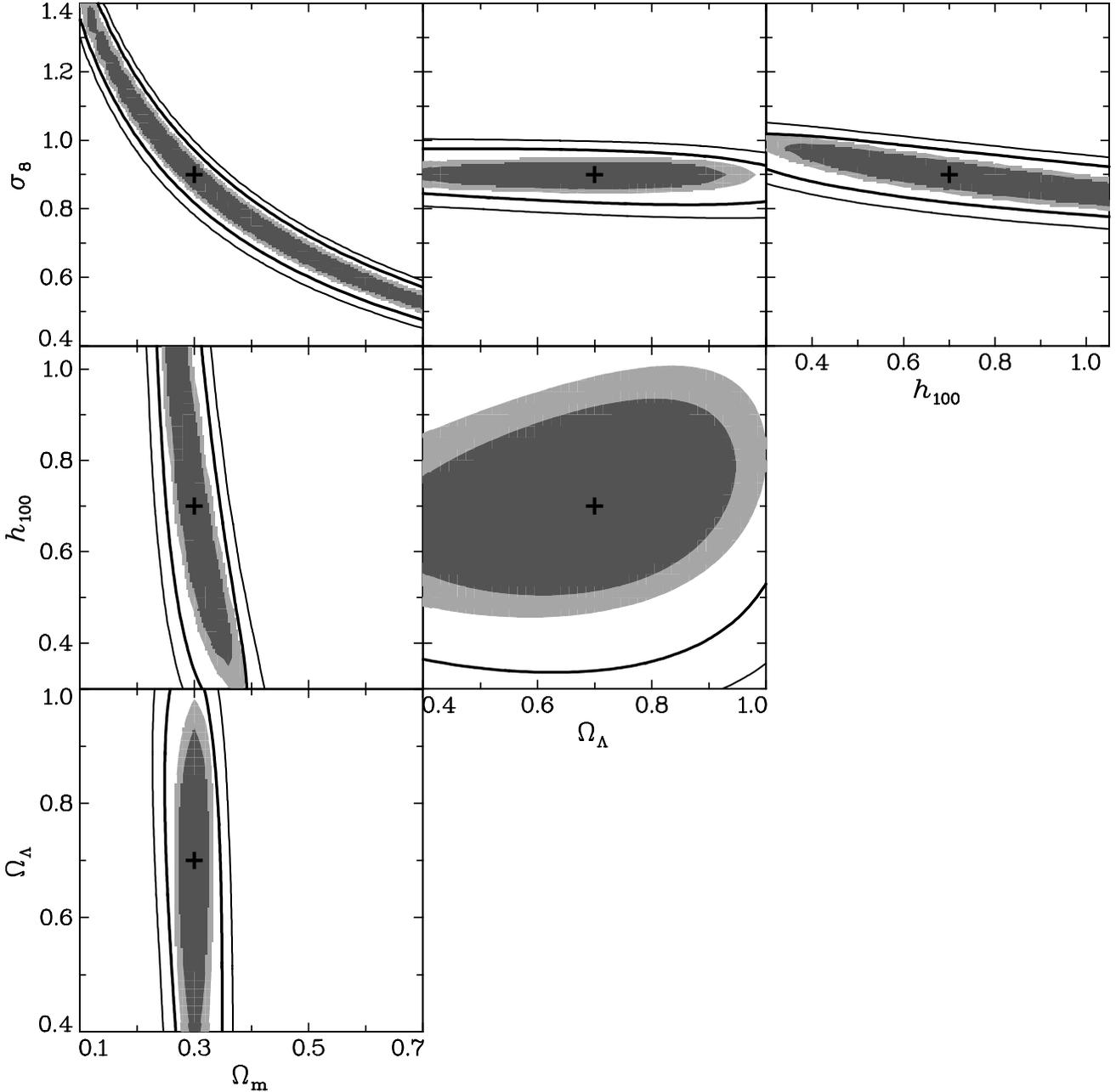}
\caption{Contours of posterior likelihood before and after shear-ellipticity nulling. Shown are all possible two-dimensional cuts through parameter space. Note that we used a fiducial survey size of $1\,{\rm deg}^2$. The cosmological parameters that are not given on the axes are evaluated at their fiducial values. In each panel the cross indicates the fiducial set of parameters. The results for the data vector $\vek{D}$, i.e. the set of tomography power spectra before nulling, are given as shaded contours, where the dark-gray area contains $60\,\%$ and the light-gray area $80\,\%$ of the posterior likelihood. The contour lines indicate the corresponding areas after shear-ellipticity nulling, using the full set, i.e. the maximum number of uncontaminated components in $\vek{D'}$. Thick lines correspond to the $60\,\%$ and thin lines to the $80\,\%$ credible region.}
\label{fig:confcontgencuts}
\end{figure*}

\begin{figure*}[t]
\centering
\includegraphics[scale=.9,angle=270]{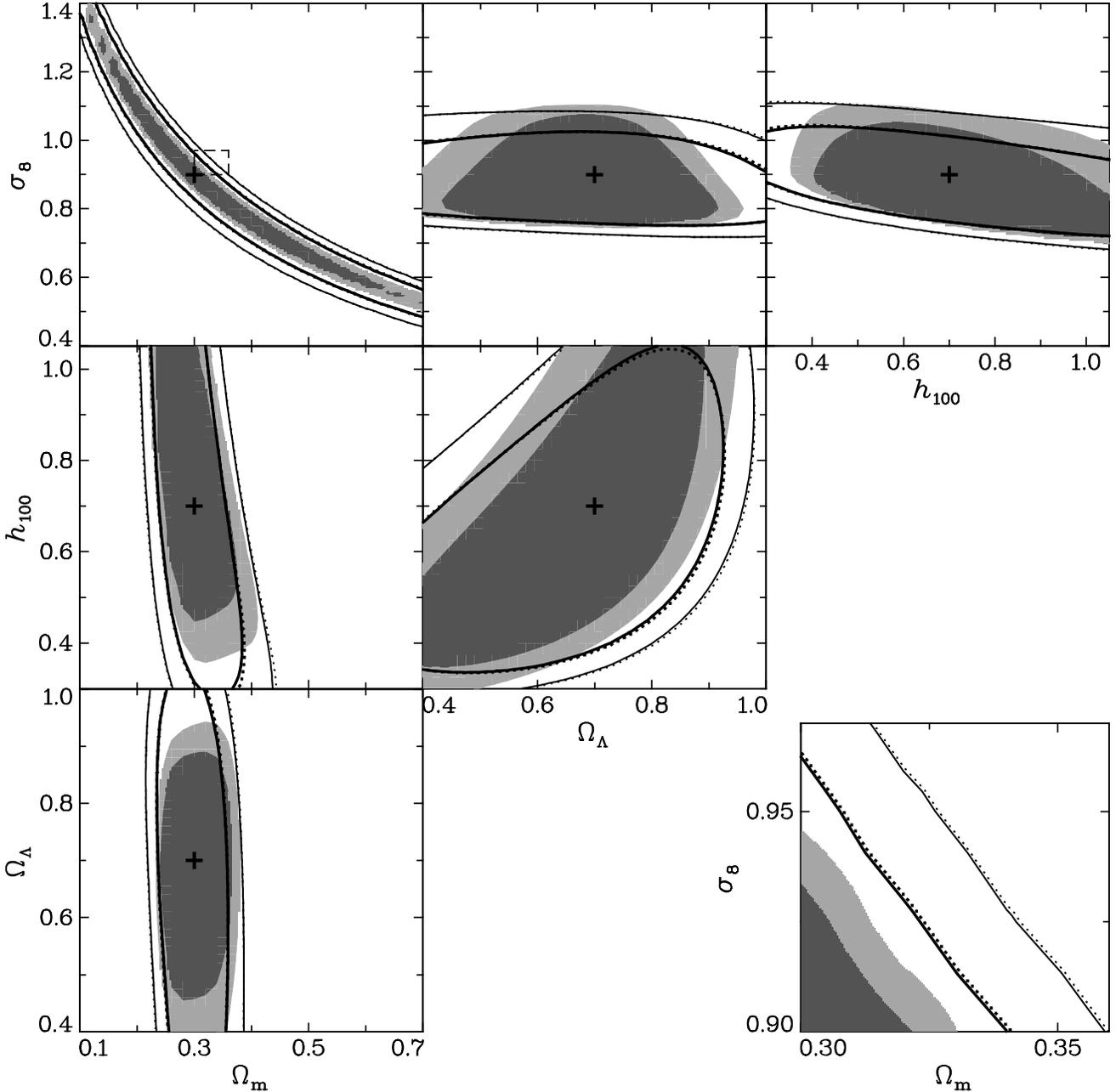}
\caption{Contours of posterior likelihood before and after shear-ellipticity nulling. Shown are the contours for all combinations of cosmological parameters out of the set $\br{\Omega_{\rm m},\Omega_\Lambda,\sigma_8,h_{100}}$, the remaining two parameters being marginalized over. The coding of areas and curves is the same as in Fig.$\,$\ref{fig:confcontgencuts}. In addition, the dotted curves enclose the credible regions resulting from using only the single optimized weight function as determined in Sect.$\,$\ref{sec:results}, i.e. power spectra of the form $\Pi^{(i)}_{[1]}(\ell)$, in $\vek{D'}$. As before, thick lines correspond to the $60\,\%$ and thin lines to the $80\,\%$ credible region, while in each panel the cross indicates the fiducial pair of parameters. The lower right panel is a detail of the upper left diagram, as outlined by the dashed box. As can be seen here, solid and dotted curves nearly coincide.}
\label{fig:confcontgenfirst}
\end{figure*}

Let the complete data vectors forming the basis of this analysis be $\vek{D}$ for the original set and $\vek{D'}$ for the transformed one, which can be written in the convenient form
\eqa{
\nn
\vek{D} &=& \Bigl( P_\kappa^{(13)}(\ell_1), ~..~ , P_\kappa^{(1 N_z)}(\ell_1), P_\kappa^{(24)}(\ell_1),  ~..~ , P_\kappa^{(2 N_z)}(\ell_1),\\ \nn
&&  \hspace*{-.3cm} P_\kappa^{(35)}(\ell_1), ~~~..~~~ , P_\kappa^{(N_z-3\, N_z-1)}(\ell_1), P_\kappa^{(N_z-3\, N_z)}(\ell_1), P_\kappa^{(N_z-2\, N_z)}(\ell_1),\\ \nn
&&  \hspace*{.5cm} P_\kappa^{(13)}(\ell_2), ~~~~~..~~~~~, P_\kappa^{(13)}(\ell_{N_\ell}),  ~~~..~~~ , P_\kappa^{(N_z-2\, N_z)}(\ell_{N_\ell}) \Bigr)\;;\\  \nn
\vek{D'} &=& \Bigl( \Pi^{(1)}_{[1]}(\ell_1), ~..~ , \Pi^{(1)}_{[N_z-2]}(\ell_1), \Pi^{(2)}_{[1]}(\ell_1), ~..~ , \Pi^{(2)}_{[N_z-3]}(\ell_1),\\ \nn
&& \hspace*{-.3cm} \Pi^{(3)}_{[1]}(\ell_1), ~~~..~~~, \Pi^{(N_z-3)}_{[1]}(\ell_1), \Pi^{(N_z-3)}_{[2]}(\ell_1), \Pi^{(N_z-2)}_{[1]}(\ell_1),\\ 
\label{eq:datavectors}
&& \hspace*{.5cm} \Pi^{(1)}_{[1]}(\ell_2),~~~~~..~~~~~, \Pi^{(1)}_{[1]}(\ell_{N_\ell}), ~~~..~~~ , \Pi^{(N_z-2)}_{[1]}(\ell_{N_\ell}) \Bigr)\;.
}
We refer to $\vek{D}$ as containing the full information although the vector is not composed of all tomography power spectra; however, as already discussed above, these entries would most probably have to be discarded anyway to avoid intrinsic ellipticity correlations. With the choice (\ref{eq:datavectors}), both vectors have the same dimension $N_{\rm D}=N_\ell (N_z-1) (N_z-2)/2$. Their components are ordered such that the corresponding covariance matrices obtain a block-diagonal structure because the power spectra evaluated at different angular frequencies are not correlated due to the assumption of Gaussianity. Hence, the covariance $C_{\rm D}$ of the data vector $\vek{D}$ reads
\eq{
\label{eq:covD}
C_{\rm D} = \br{ \begin{array}{cccc} \ba{\Delta \vek{d}_1 ~\Delta \vek{d}^\tau_1} & 0 & ... & 0\\ 0 & \ba{\Delta \vek{d}_2 ~\Delta \vek{d}^\tau_2} & ... & 0\\ ... & ... & ... & 0\\ 0 & 0 & 0 & \ba{\Delta \vek{d}_{N_\ell} ~\Delta \vek{d}^\tau_{N_\ell}} \end{array} }\;,
}
where for the sake of a compact notation, the vector
\eq{
\vek{d}_i := \br{P_\kappa^{(13)}(\ell_i), ~..~ , P_\kappa^{(1 N_z)}(\ell_i), P_\kappa^{(24)}(\ell_i), ~~~..~~~ , P_\kappa^{(N_z-2\, N_z)}(\ell_i)} 
}
was introduced, so that $\vek{D} = \br{\vek{d}_1, ~..~ , \vek{d}_{N_\ell}}$. The remaining non-trivial blocks $\ba{\Delta \vek{d}_i ~\Delta \vek{d}^\tau_i}$ for the angular frequency bin $\ell_i$ with dimension $(N_z-1) (N_z-2)/2 \times (N_z-1) (N_z-2)/2$ each are computed by means of (\ref{eq:covPtom}) and can then readily be inverted numerically. The covariance of $\vek{D'}$ is dealt with analogously.

Since the set of $\Pi^{(i)}_{[q]}(\ell)$ with $q=1,\,..\,,N_z-i-1$ contains the full information, the optimization of the weight functions with respect to the trace of the Fisher matrix becomes superfluous in this situation. Instead, one can construct the $\vek{B}^{[q]}$ simply as a set of orthogonal vectors, starting with $\vek{f}$, for instance by means of the Gram-Schmidt algorithm. This way the vectors corresponding to the weight functions still fulfill the constraint equation (\ref{eq:anaconstraint}).

\begin{table}[h]
\begin{minipage}[t]{\columnwidth}
\renewcommand{\footnoterule}{}  
\caption{Set of cosmological parameters used for the analysis.}
\begin{tabular}[t]{c|c|c|c|c|c}
& $p_{\rm f}$ & $p_{\rm min}$ & $p_{\rm max}$ & $p_{\rm l}$ & $p_{\rm u}$\footnote{Note: values of the fiducial model ($p_{\rm f}$), limiting values $p_{\rm min}$ and $p_{\rm max}$ of the parameter plane considered, and lower ($p_{\rm l}$) and upper limits ($p_{\rm u}$) of the prior applied in the marginalization.}\\
\hline\hline
$\Omega_{\rm m}$ & 0.3 & 0.10 & 0.70 & 0.20 & 0.40\\
$\Omega_\Lambda$ & 0.7 & 0.40 & 1.00 & 0.58 & 0.82\\
$\sigma_8$ & 0.9 & 0.40 & 1.40 & 0.80 & 1.00\\
$h_{100}$ & 0.7 & 0.30 & 1.05 & 0.60 & 0.80\\
\end{tabular}
\label{tab:parameters}
\end{minipage}
\end{table}

To calculate credible regions, the likelihood function in parameter space has to be evaluated, which reads under the assumption of a Gaussian probability distribution function
\eq{
\label{eq:likelihood}
L(\vek{D}|\vek{p}) = \frac{1}{(2\pi)^\frac{N_D}{2} \sqrt{\det C_D}} ~\expo{ -\frac{1}{2} \bb{\svek{D}(\svek{p})-\svek{D}_{\rm f}}^\tau {C_D}^{-1} \bb{\svek{D}(\svek{p})-\svek{D}_{\rm f}}}
}
and likewise for $\vek{D'}$, where $\vek{D}_{\rm f}$ stands for the data vector, as obtained for the fiducial model, and $\vek{p}$ again denotes the set of varied cosmological parameters. The covariance matrices are only evaluated at the fiducial cosmology, as well as the set of weight functions $B^{(i)}_{[q]}(\chi)$ entering the measures in $\vek{D'}$. We assume flat priors on the whole range of parameters considered, leading to a posterior likelihood
\eq{
\label{eq:likepost}
L_{\rm post}(\vek{p}|\vek{D}) = \frac{L(\vek{D}|\vek{p})}{\sum_{\vek{p}} L(\vek{D}|\vek{p})}\;.
}
The boundaries of the four-dimensional grid in parameter space over which the sum in the equation above runs and the fiducial cosmological parameters are summarized in Table$\,$\ref{tab:parameters}. Otherwise the setup described in Sect.$\,$\ref{sec:results} is kept, except for a number of adjustments owing to the restrictions in computational power. The power spectra are now calculated for $N_\ell=30$ bins in the range between $\ell=50$ and $\ell=10^4$, making use of the fit formula for the non-linear structure evolution by \citet{PeacockDodds}. In addition, the number of redshift bins is reduced to $N_z=20$, still ranging from $z=0$ to $z=4$.

\begin{figure*}[t]
\centering
\includegraphics[scale=.6,angle=270]{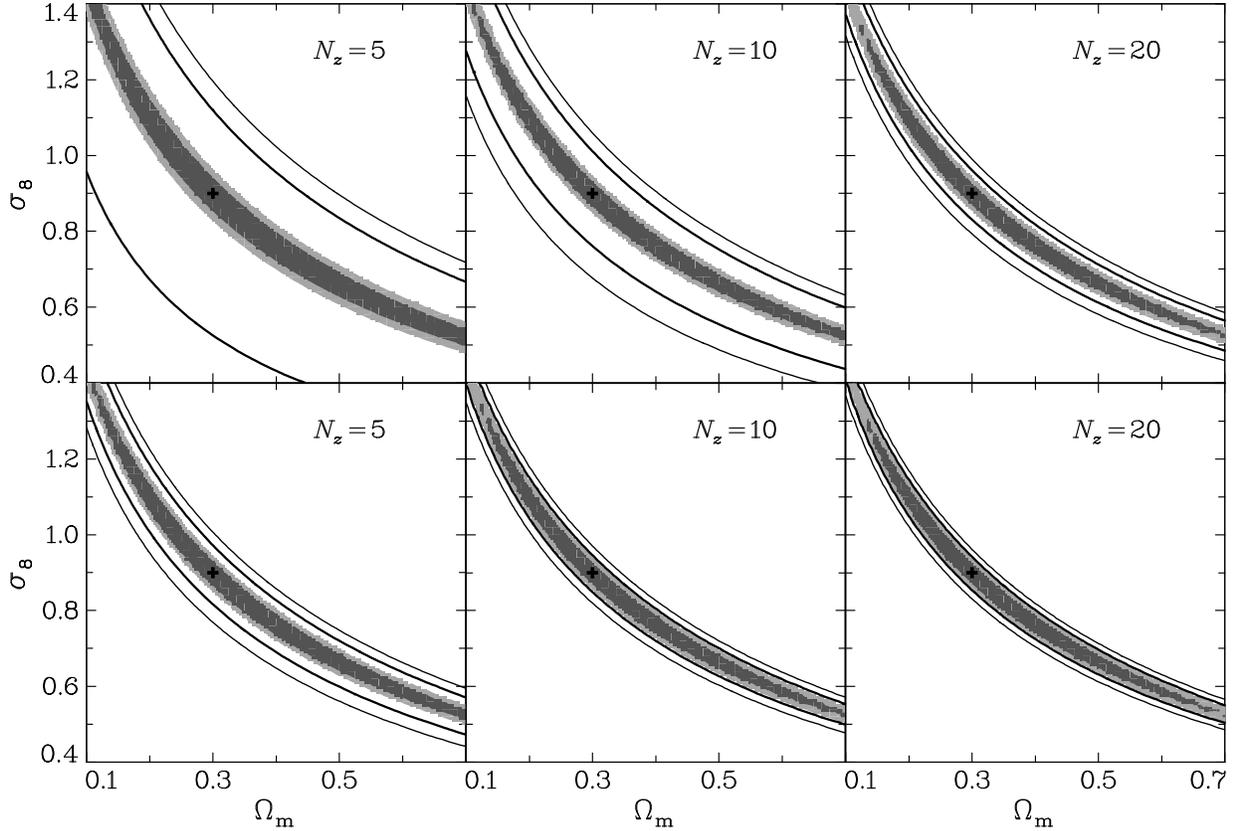}
\caption{Credible regions in the $\Omega_{\rm m}-\sigma_8$ plane for varying redshift binning before and after shear-ellipticity nulling, using the full set of measures. In the upper panels the data vectors as given in (\ref{eq:datavectors}) are used; for the results shown in the lower panels the cross-correlation power spectra of adjacent bins were also incorporated into both $\vek{D}$ and $\vek{D'}$. The data sets used are otherwise identical, except for the number of redshift bins, which is $N_z=5$ in the left panels, $N_z=10$ in the central panels, and $N_z=20$ in the right panels. The coding of areas and curves is the same as in Fig.$\,$\ref{fig:confcontgencuts}. Again, the crosses indicate the fiducial set of cosmological parameters.}
\label{fig:confcontgen2D}
\end{figure*}

If all entries in $\vek{D'}$ of the form $\Pi^{(i)}_{[N_z-i-1]}(\ell)$ are removed, only components free of shear-ellipticity correlations remain. The loss of information caused by this removal is illustrated in Fig.$\,$\ref{fig:confcontgencuts} where two-dimensional cuts through the credible regions in parameter space, resulting before and after nulling, are given. The parameters not shown are evaluated at their fiducial values, so that the cross in each panel marks the fiducial model, in this case coinciding with the point of maximum likelihood. In the $\Omega_{\rm m}-\sigma_8$ plane, one recognizes the typical banana shape, while the Hubble parameter and the density of dark energy are only poorly constrained by our setup. As expected, the contours after the application of the nulling technique have widened throughout. The inner contour line remains outside the light-gray area, implying that the probability that a range of parameters contains the true cosmological model decreases from $80\,\%$ to less than $60\,\%$ after the removal of contamination by intrinsic alignment. It is interesting to note that the ratio of the $\chi^2$, i.e. the argument of the exponential in (\ref{eq:likelihood}), before and after nulling is roughly constant over the whole range of parameters considered.

In Fig.$\,$\ref{fig:confcontgenfirst} the same set of credible regions is plotted, but here the hidden parameters have been marginalized over with flat priors within the range indicated in Table$\,$\ref{tab:parameters}. The appearance of the contours in the $\Omega_{\rm m}-\sigma_8$ plane remains similar due to the small influence of $h_{100}$ and $\Omega_\Lambda$. However, the characteristic banana-like shape causes the area of maximum likelihood to be shifted away from the fiducial model in panels where one of the parameters $\Omega_{\rm m}$ or $\sigma_8$ is marginalized over. Consequently, the posterior likelihood peaks at lower values of $\sigma_8$ in the top center and right panel, while $L_{\rm post}$ obtains its maximum at lower values than $\Omega_{\rm m}=0.3$ in the central left and lower left panel. Besides that, the nulling implies that the degeneracy in $h_{100}$ and $\Omega_\Lambda$ even increases, see Fig.$\,$\ref{fig:confcontgencuts}, which leads to a significant elongation of the credible regions along these parameters in the panels of Fig.$\,$\ref{fig:confcontgenfirst} mentioned above. Although this stretching causes lines of equal likelihood to even intersect, thereby apparently improving parameter constraints in one dimension, it is expected that the total area of a credible region increases by shear-ellipticity nulling, which corresponds to an overall decrease in constraints. This is indeed the case as will be shown below.

In addition, we have computed the posterior likelihood for nulling with first-order measures alone, i.e. for a vector $\vek{D'}$ with all components apart from those with a subscript $[1]$ removed. The resulting contours are also presented in Fig.$\,$\ref{fig:confcontgenfirst}. Within the resolution of the graphics and the grid we employed to cover the parameter space, the contours for the first-order nulling coincide with the results for the full set\footnote{Note that \lq full set\rq\ means the use of the maximum number of new power spectra that are not contaminated.}. Only for the more concentrated likelihood in the $\Omega_{\rm m}-\sigma_8$ plane are the first-order contours located distinctly farther outside, as can be seen in the inlet of Fig.$\,$\ref{fig:confcontgenfirst}. The values where the contours are drawn for the two setups deviate by less than $3\,\%$ in the latter case; for all other parameter planes, the deviation of the contour values is approximately $1\,\%$.

We quantify the widening of contours in terms of the quadrupole moments of the likelihood function, employing $q$-values as introduced by \citet{kilbinger04}. For any two-dimensional likelihood analysis, one defines
\eq{
\label{eq:quadrupole}
Q_{\mu\nu} = \sum_{\vek{p}} L_{\rm post}(\vek{D}|\vek{p}) \br{p_\mu - p_{{\rm f},\mu}} \bigl( p_\nu - p_{{\rm f},\nu} \bigr) 
}
for $\mu,\nu=1,2$. Then the quantity
\eq{
\label{eq:qvalue}
q = \sqrt{\det Q} = \sqrt{Q_{11}~Q_{22} - Q_{12}^2}
}
scales with the area of the credible region. Hence, an increase in $q$-value corresponds to a degradation of parameter constraints. We compute the relative change in $q$ due to nulling, using both the full set and only first-order measures. The results, listed in Table$\,$\ref{tab:qmarg}, confirm the only marginally weaker performance of the first-order-only configuration. Depending on the combination of parameters, the increase in $q$ ranges between $20\,\%$ and about $50\,\%$.

\begin{table}[h]
\begin{minipage}[t]{\columnwidth}
\renewcommand{\footnoterule}{}  
\caption{Increase in $q$-values for the marginalized credible regions shown in Fig.$\,$\ref{fig:confcontgenfirst}.}
\begin{tabular}[t]{c|c|c|c}
$p_1$ & $p_2$ & $\Delta q_{\rm tot}$ [\%]& $\Delta q_1$ [\%]\footnote{The relative change in $q$ is given for nulling with the full set $\br{\Delta q_{\rm tot}}$ and with first-order measures only $\br{\Delta q_1}$.}\\
\hline\hline
$\Omega_{\rm m}$ & $\sigma_8$ & 29.2 & 30.3\\
$\Omega_{\rm m}$ & $h_{100}$ & 45.6 & 48.3\\
$\Omega_{\rm m}$ & $\Omega_\Lambda$ & 50.3 & 52.5\\
$\Omega_\Lambda$ & $\sigma_8$ & 36.4 & 37.5\\
$\Omega_\Lambda$ & $h_{100}$ & 21.2 & 21.7\\
$h_{100}$ & $\sigma_8$ & 26.4 & 27.3\\
\end{tabular}
\label{tab:qmarg}
\end{minipage}
\end{table}

\begin{table}[h]
\begin{minipage}[t]{\columnwidth}
\renewcommand{\footnoterule}{}  
\caption{Increase in $q$-values for credible regions, resulting from different redshift binning of the survey.}
\begin{tabular}[t]{c|c|c}
$N_z$ & $\Delta q$ [\%] & $\Delta q'$ [\%]\footnote{The values $\Delta q$ correspond to the regions shown in the upper panels, the values $\Delta q'$ to the regions in the lower panels of Fig.$\,$\ref{fig:confcontgen2D}.}\\
\hline\hline
5 & 192.5 & 35.5\\
10 & 92.7 & 17.2\\
20 & 28.9 & 15.6\\
\end{tabular}
\label{tab:qnzbins}
\end{minipage}
\end{table}

To assess the importance of a large number of redshift bins, we considered a further setup with only two varied parameters $\br{\Omega_{\rm m},\sigma_8}$, but otherwise identical with respect to the foregoing implementation. We used 5, 10, and 20 redshift bins to cover the range between $z=0$ and $z=4$. The resulting credible regions before and after nulling, using the full set of measures, are given in Fig.$\,$\ref{fig:confcontgen2D}, upper panels. Concerning the original set of tomography power spectra, the increase in the number of redshift bins does not improve the parameter constraints appreciably, as already discussed for instance by \citet{hu99}. In contrast to this, the loss of information due to the removal of potentially contaminated components of the data vector is dramatic for $N_z=5$ and still considerable for $N_z=10$; see Table$\,$\ref{tab:qnzbins} for the corresponding changes in $q$-values. In the lower panels of Fig.$\,$\ref{fig:confcontgen2D} the credible regions, resulting from the inclusion of the cross-correlation power spectra of adjacent bins into the data vectors (\ref{eq:datavectors}), are shown (see also Table$\,$\ref{tab:qnzbins}). The contours tighten substantially, in particular in the case $N_z=5$, where the probability of a contamination by intrinsic alignment of cross-correlations between neighboring bins is low anyway due to the large bin size. With this setup, a number of 10 redshift bins already ensures that parameters are still well-constrained after nulling.

\section{Conclusions}
\label{sec:conclusions}

In this paper we have elaborated on a purely geometrical method to eliminate shear-ellipticity correlations, which constitute a source of severe contamination of the cosmic shear signal. Using a nulling technique, new observables free of this contamination are constructed by suitably weighting shear tomography power spectra. The weighting is determined such that the contribution to the cosmic shear signal from matter structures potentially causing shear-ellipticity correlations is removed by means of the characteristic dependence of these correlations on redshift.

Three approaches to obtaining weight functions were investigated, which in addition optimize the information content of the new observables in terms of the Fisher matrix. The results for both analytical and numerical methods are in good agreement, also for higher orders that are constructed by imposing a suitably defined condition of orthogonality. Most notably, the analytical ansatz, being computationally simple, is compatible in spite of several simplifications.

Using a set of tomographic power spectra with 20 redshift bins, we computed credible regions in a four-dimensional parameter space to assess the loss of information due to shear-ellipticity nulling. The contours widen significantly with an increase in $q$-values of up to $50\,\%$ although reasonably stringent constraints on cosmological parameters are still possible. Besides, it was demonstrated that the use of a smaller subset of power spectra, excluding those constructed by means of higher-order weight functions, yields a practically identical performance, compared to the full set.

However, data with a smaller number of redshift bins could be shown to almost completely lose its ability to constrain cosmological parameters when intrinsic alignment is taken into account (see also BK07), unless the contamination by intrinsic ellipticity correlations of power spectra cross-correlating adjacent bins is safely under control. Hence, a large number of redshift bins is desirable for cosmic shear studies, in spite of the fact that, without the effects of intrinsic galaxy alignment, constraints on cosmological parameters are not appreciably improved once $N_z \gtrsim 5$. These results underline once more the need for both precise and detailed redshift information to control systematics in cosmic shear.

We emphasize that, using the full set of new power spectra, a maximization of the information content in the weight functions is not necessary. In particular, this means that shear-ellipticity nulling as such does not rely on the Fisher matrix formalism and the determination of optimal weight functions, nor do the credible regions that result from applying the nulling. If only the first-order power spectra are employed, parameter constraints do depend on the optimization of weight functions. However, the first-order results agree well for the three considered approaches, so that the change of the likelihood contours due to the use of these different methods is expected to be only marginal.

The moderate change in $q$-values indicates that this method is in principle suited to inferring cosmological parameters with fair precision. It is currently the only truly secure means of safely eliminating shear-ellipticity -- and in addition, by construction, intrinsic ellipticity -- correlations since no assumptions are made about the still uncertain models of these systematic effects. An improvement in the performance of shear-ellipticity nulling could be achieved by taking advantage of the intrinsic ellipticity correlations not being only restricted to galaxy pairs which are close in redshift, but also close on the sky, thereby allowing for keeping part of the signal from correlations of adjacent redshift bins.

As soon as reliable data exists for modeling the intrinsic alignment of galaxies, an intermediate approach between strict shear-ellipticity nulling and the full reliance on the shear-ellipticity power spectrum could be developed. Since the tidal forces acting on a galaxy are caused by the surrounding matter-density distribution, correlations between matter density and intrinsic ellipticity can be used to determine the tidal field and, consequently, the expectation value of the orientation of the intrinsic ellipticity. This additional information, appropriately incorporated into the nulling technique, may improve parameter constraints, while keeping the influence of model uncertainties at a low level. \citet{mandelbaum06} have analyzed density-ellipticity correlations in the Sloan Digital Sky Survey, using galaxies as tracers for the mass; see references therein for further observations.

A central aspect to be considered before applying the nulling technique to real data is the influence of photometric redshift uncertainty since this will further deteriorate constraints (see e.g. BK07, \citealp{abdalla07}) and possibly enforce a modification of the method. Catastrophic outliers in the determination of photometric redshifts may cause a leakage of shear-ellipticity correlations into the measures constructed by the nulling technique. Currently, we are not able to judge how severely this leakage will affect our method. In a follow-up study, we will investigate the demands on photometric redshift accuracy due to nulling in detail. However, the probability distribution of galaxy distances enters our method only via the shape noise term in the power spectrum covariance matrix, see (\ref{eq:covPtom}), which is merely used for determining optimized weight functions and not needed for constructing the full set of measures, suggesting that shear-ellipticity nulling could be fairly robust against incomplete knowledge of the distance distribution. Note that, even if very accurate photometry were available, the determination of distance in terms of redshift would still be limited by the peculiar velocities of the observed galaxies and therefore provide an upper limit to the number of usable redshift bins.

Furthermore, it would be of interest to determine the decrease in accuracy due to nulling, when determining the equation-of-state parameters $w_0$ and $w_a$ of dark energy, because upcoming observations will set a focus on these parameters and, besides, the results would allow for a direct comparison with other methods whose performance is mainly evaluated in terms of the dark energy figure of merit. In addition, one may think of generalizations such as the application to higher-order statistical measures of cosmic shear. As far as third-order statistics are concerned, neither shear-ellipticity nor intrinsic ellipticity correlations have been subject to much investigation so far. Recently, \citet{semboloni08} have used simulations to demonstrate that the three-point shear-ellipticity correlation terms can reach up to $10\,\%$ of the amplitude of the pure shear signal in a survey with a median redshift of 0.7.

Irrespective of the future practical relevance of the approach presented in this study or its possible modifications, the limited abilities of this geometrical approach to put constraints on cosmological parameters prove the necessity of improving the understanding of galaxy formation and evolution in order to develop accurate models of intrinsic galaxy alignment. Note that, if a good model were known someday, it would be possible to even tighten constraints on cosmological parameters since shear-ellipticity correlations are another cosmological probe complementary to the pure lensing signal. The detailed knowledge of intrinsic alignment would not only allow for the efficient removal of contaminations of the cosmic shear signal, but in addition provide an interesting means to study the interaction between galaxies and their environment.

\begin{acknowledgements}
We would like to thank Jan Hartlap for providing the code for calculating power spectra. Furthermore, we are grateful to our referee for the comments that substantially improved this article. BJ acknowledges support by the Deutsche Telekom Stiftung. This work was supported by the Deutsche Forschungsgemeinschaft under the projects SCHN 342/6--1, the Priority Program 1177 `Galaxy Evolution', and the Transregional Collaborative Research Center TR33 `The Dark Universe'.
\end{acknowledgements}

\bibliographystyle{aa}

\end{document}